\documentclass[preprint,3p,12pt]{elsarticle}

\usepackage{graphicx}%
\usepackage{multirow}%
\usepackage{amsmath,amssymb,amsfonts}%
\usepackage{amsthm}%
\usepackage{mathrsfs}%
\usepackage[title]{appendix}%
\usepackage{xcolor}%
\usepackage{textcomp}%
\usepackage{manyfoot}%
\usepackage{booktabs}%
\usepackage{algorithm}%
\usepackage{algorithmicx}%
\usepackage{algpseudocode}%
\usepackage{listings}%
\usepackage{subcaption}
\DeclareMathOperator{\e}{e}

\theoremstyle{thmstyleone}%

\theoremstyle{thmstyletwo}%

\theoremstyle{thmstylethree}%
\usepackage{float}

\begin{document}

\begin{frontmatter}



\title{A model-inversion control strategy to attain desired equilibrium points when synthesizing nonlinear resonators}


\author[TPE,ECL2]{Maxime Morell} 
\author[ECL2]{Emanuele De Bono} 
\author[TPE]{Emmanuel Gourdon} 
\author[ECL]{Manuel Collet} 
\author[TPE]{Alireza Ture Savadkoohi} 
\author[TPE]{Claude-Henri Lamarque} 
\affiliation[TPE]{organization={ENTPE, Ecole Centrale de Lyon, CNRS, LTDS, UMR5513},
            addressline={3 rue Maurice Audin}, 
            city={Vaulx-en-Velin},
            postcode={69518}, 
            state={},
            country={France}}
            
\affiliation[ECL2]{organization={Ecole Centrale de Lyon, CNRS, ENTPE, LTDS, UMR5513},
            	addressline={36 av. Guy de Collongue}, 
            	city={Ecully},
            	postcode={69130}, 
            	state={},
            	country={France}}
    
\affiliation[ECL]{organization={CNRS, Ecole Centrale de Lyon, ENTPE, LTDS, UMR5513},
	addressline={36 av. Guy de Collongue}, 
	city={Ecully},
	postcode={69130}, 
	state={},
	country={France}}

\begin{abstract}
The initial conditions of a deterministic nonlinear system affects the attained asymptotic solutions. This is a problem when the nonlinear behaviour is achieved by active control synthesis, and the initial conditions are not accessible for pre-constraining. Meanwhile, the generalized impedance control is recently gaining increasing attention in acoustics, and has been enlarged to achieve nonlinear mechanical responses of Electroacoustic Resonators at low excitation levels. Nevertheless, as the initial conditions of Electroacoustic Resonators are hardly accessible, the equilibrium points actually showcased by the synthetic nonlinear dynamics are not advantageous for noise attenuation in cavities excited by realistic noise sources. In this paper, we propose a control strategy being able to attain the desired equilibrium points, due to the definition of an exosystem characterized by a desired nonlinear impedance and a synthetic external excitation. The exosystem dynamics is integrated in real time and its response is enforced in the actual system (in our case, the Electroacoustic Resonator), by a model-inversion feedforward strategy. The proposed algorithm is implemented in the Electroacoustic Resonator, and is validated both numerically and experimentally. In particular, the experimental testing is conducted both in a quasi-open acoustic environment, and in enclosed cavities for acoustic mode attenuation.
\end{abstract}


\begin{highlights}
\item The model-inversion method is modified to target specific equilibrium points
\item The use of short transient synthetic force is used in a numerical ExoSystem
\item The control strategy is applied to an electroacoustic resonator
\item The method is validated through numeric basins of attractions and two experiments
\end{highlights}


\end{frontmatter}


\section{Introduction}

Nonlinear resonators have received increasing attention in vibroacoustics, from the early works of \cite{roberson1952synthesis,jordanov1988optimal}, up to the conception of the Nonlinear Energy Sink \cite{vakakis2001inducing} for achieving the so-called Targeted Energy Transfer where vibrational energy is transferred from the linear host structure to the nonlinear absorber \cite{aubry2001analytic}, \cite{vakakis2008nonlinear}. More recently, the potential applications of nonlinear resonators have been investigated also in acoustics on Helmholtz resonators \cite{alamo2018nonlinear} or membranes \cite{cochelin2006experimental,bellet2010experimental,bouzid2025towards}. In both cases, the nonlinear behaviour of the resonator requires high excitation levels to be triggered. Moreover, in \cite{cochelin2006experimental,bellet2010experimental,bouzid2025towards} a pre-stress is required on the membrane for achieving the desired nonlinear response. To improve robustness, Electromechanical coupling and active control have been recently combined with passive nonlinear absorbers for the vibration control of solid structures \cite{mesny2024nonlinear,rodriguez2024sliding}. Meanwhile, the digital synthesis of acoustic impedances by so-called Electroacoustic Resonators (ER) has quickly evolved, from linear and local tunable impedances \cite{rivet2016broadband,de2022effect,billon2022flow}, towards the implementation of so-called \emph{generalized} impedances, as described in the comprehensive review \cite{mallejac2025active}. The classical local and linear impedance concept can indeed be generalized to include nonlocal \cite{DeBono2024j}, or nonlinear behaviors \cite{guo2020improving,DeBono2024,MORELL2024118437,morell2025experimental}. In particular, the advantages of the digital synthesis of nonlinear generalized impedance is the possibility to showcase nonlinear behaviours at low excitation levels, and to easily tune the nonlinear resonator's parameters. Moreover, while in \cite{guo2020improving} the Duffing-like ER response was achieved by simply cubing the estimated ER displacement available in feedback, in \cite{DeBono2024,MORELL2024118437,morell2025experimental} a more sophisticated model-inversion feed-forward strategy (called Real-Time-Integration (RTI) algorithm) has been implemented which allows to exhibit unprecedented non-polynomial nonlinear behaviours. However, in all the above synthetic nonlinear systems, the selection of the ER response, among the multiple stable equilibrium points, have been left to the external environment and excitation, as initial conditions are uncontrolled and pre-constrains are unfeasible. The first and main novelty of the present paper is the enlargement of the RTI model-inversion strategy to target specific equilibrium points of the synthetic nonlinear resonator, for any initial condition. The RTI concept belongs to the class of feedforward model-inversion based approaches which are classically employed to supply feedback loops for the tracking of either linear or nonlinear exosystems (ESs) through the so-called regulator theory \cite{Francis1976,Isidori1990,devasia1996nonlinear}. In the present paper, though, the RTI algorithm is conceived as purely feedforward, but our approach might be easily enlarged to contemplate feedback loops by exploiting different ER architectures as the one employed in \cite{guo2020improving}. The definition of an Exosystem (ES) in a Ordinary Differential Equation (ODE) form, characterized by nonlinear dynamics under artificial excitations, is what allows our RTI algorithm to target specific equilibrium points, and is for the first time implemented and validated in this paper.\\
The paper is structured as following: Section \ref{sec1} provides the general mathematical framework of the RTI model-inversion control, while Section \ref{sec:targeting a specific equilibrium point} introduces the strategy to target specific equilibrium points. Then, Section \ref{sec:the case of ER} particularizes the strategy for the ER application, and Section \ref{sec:numerical simulations} provides its numerical validation. Finally, Sections \ref{sec:open field experiment} and \ref{sec:coupled} validate the concept in two experimental environments: a quasi-open acoustic field, and a small cavity, respectively. Conclusions and next steps are given in Section \ref{sec:conclusions}.

\section{Nonlinear impedance control and basic assumptions}\label{sec1}

We seek to create a feed-forward impedance control method that allows to digitally program nonlinear behaviors for a plant to adopt, at low excitation amplitudes. \\

Let us consider the plant model for a general system, linear or nonlinear, with $n\in\mathbb{N}^*$ degrees of freedom:
\begin{equation} \label{eq:plant}
	\dot X(t)=A(X(t), t, f_{ext}(t), u_d(t))
\end{equation}
where $X\in\mathbb{R}^n$ stands for the state vector, $f_{ext}(t)\in\mathbb{R}^n$ denotes the vector of exterior forces, $u_d\in\mathbb{R}^n$ is the vector of the desired control commands, $t\in[0,T],\quad T>0$ represents the time and the upper dot ($\dot \bullet$) the time derivative. As a result, the vector function $A$ describes the evolution of the system states and is defined as $A:(\mathbb{R}^n \times [0,T] \times \mathbb{R}^n \times \mathbb{R}^n) \longrightarrow \mathbb{R}^n$. $A$ is assumed to be of class $C^k$ with $k\geq1$. \\

\subsection{The set-point}
For each plant to be controlled, a set-point is the plant state to be achieved. We aim at enforcing nonlinear behaviors into linear systems at low excitation amplitudes. The method proposed in this study may be used to target linear responses in nonlinear systems, but it will not be studied here.\\

To target nonlinear behaviors into linear systems, the method proposed in this paper consists in the introduction of an ES fed by measurements. The ES is a numerical oscillator corresponding to the desired behaviour of the plant, and governed by the equation:
\begin{equation}\label{eq:desired dynamics}
	\dot X_d (t)= F(X_d(t),t,f_{ext}(t))
\end{equation}
where $X_d \in\mathbb{R}^n$ stands for the desired vector of system states and the vector function $F:(\mathbb{R}^n \times [0,T] \times \mathbb{R}^n) \longrightarrow \mathbb{R}^n$ describes the mathematical model of the desired behaviour of the plant. The numerical system receives the vector of exterior forces $f_{ext}$, which can be taken as the measured vector of exterior forces $f_{ext}^m$. The ES is the target dynamics informed by the external forces applied on the plant. As a result, solving the ES dynamics provides the desired vector state $X_d$. Hence, the objective of the controller is to enforce the ES vector states into the plant. If the desired dynamics $F$ is nonlinear, multiple stable solutions of Eq. \eqref{eq:desired dynamics} can be featured, and the solution exhibited by the ES will depend upon the initial conditions representing as $X_d(t_0)=X_{d0}$, and $f_{ext}(t_0)=f_{ext,0}$.

\subsection{A model-inversion based feed-forward controller} \label{sec:conditions}
To enforce the desired behaviour provided by the ES, a feed-forward controller is employed suiting the available experimental architecture. 
A way to build a feed-forward controller, is to invert the plant system dynamics. Let us define the function $g$:

\begin{equation}\label{eq:g definition}
	\begin{split}
		&g(\dot X(t),X(t),t,f_{ext}(t),u_d(t))=\\
		&\dot X(t) - A(X(t),t,f_{ext}(t),u_d(t)).
	\end{split}
\end{equation}

According to the implicit function theorem \cite{Zeidler1986}, the function $g$ must respect some assumptions in order to be able to obtain the explicit form of the controller $u_d$. First of all, we can assign the Banach spaces $H=\mathbb{R}^n$ and $E= \mathbb{R}^n \times \mathbb{R}^n \times [0,T] \times \mathbb{R}^n$, respectively to $u_d$ and the remaining variables $(\dot X(t),X(t),t,f_{ext}(t))$ (grouped in $r$) of $g$. Then, we can assume that there exists $\exists (r_0, u_{d,0})\in E\times H$ such that $g(r_0, u_{d,0})=0$. So, according to the implicit function theorem \cite{Zeidler1986}, if the application $g$ is of class $C^k, \ k\geq1$, and the matrices:
\begin{equation}
	\biggr(\frac{\partial g_p}{\partial (u_{d})_q}\biggr|_{r=r_0,u_d=u_{d,0}}\biggr)_{1\leq p \leq n,\ n+1\leq q \leq 2n}
\end{equation}
are invertible, with continuous inverses, then, the neighbourhoods $E'$ of $r_0$ and $H'$ of $u_{d,0}$ exist so that:
\begin{equation}
	\begin{split}
		&\exists \varphi\in C^k(E',H')\ , \forall (r,u_d)\in E'\times H'\\
		&g(r,u_d)=0 \Leftrightarrow u_d=\varphi(r).
	\end{split}
\end{equation}
This means that, if the function $g$ is continuously differentiable, and its derivatives with respect to $u_d$ are invertible, and such inverses are continuous, then we can express explicitly $u_d$ as function of the remaining variables $r$: $u_d=\varphi(r)$. In that case, the expression $u_d=\varphi(r)$ can be exploited to synthesize a feed-forward controller, according to the so-called model-inversion method. Observe that the above theorem is valid for either linear or nonlinear plants: a model-inversion based feed-forward controller can be synthesized as long as the actuation variable respects the above assumptions. 

Let us suppose that the setpoint $X_d$ is known (by integration of Eq. \eqref{eq:desired dynamics}), the controller $u_d$ being able to track such setpoint in the plant, is obtained by imposing $\forall t\in [0,T]$ $X(t)=X_d(t)$ in the plant Eq. \eqref{eq:plant}. Therefore:
\begin{equation}
	\begin{split}
		&\forall (X_d,t,f_{ext},u_d)\in E'\times H'\\
		&\dot X_d(t)=A(X_d(t), t, f_{ext}^m(t), u_d(t))\\ 
		\Leftrightarrow \; &u_d=\varphi(\dot X_d(t),X_d(t),t,f_{ext}^m(t)).
	\end{split}
\end{equation}

This imposes the additional condition that $X_d$ satisfies the equation of the plant Eq. (\ref{eq:plant}) $\forall t\in [0,T]$.

\subsection{General formulation}\label{sec:general formulation}
Based on the described procedures in previous sections, under the assumptions of the implicit function theorem, the entire controlled system (the plant and the controller) is described by the following system of equations:
\begin{subequations}\label{eq:global}
	\begin{equation}\label{eq:plant_g}
		\dot X(t)=A(X(t), t, f_{ext}(t), u_d(t))
	\end{equation}
	\begin{equation}\label{eq:ES}
		\dot X_d (t)= F(X_d(t),t,f_{ext}^m(t)).
	\end{equation}
	\begin{equation}\label{eq:ud}
		u_d(t)=\varphi(\dot X_d(t), X_d(t),t,f_{ext}^m(t)).
	\end{equation}
\end{subequations}
To describe this controller, we assumed that $\forall t\in [0,T] \quad X(t)=X_d(t)$, which is also the aim of the controller.
Indeed, if Eq. \eqref{eq:ud} is obtained from Eq. \eqref{eq:plant_g}, by model-inversion (see Section \ref{sec:conditions}) and supposing $f_{ext}^m\approx f_{ext}$, since Eq. \eqref{eq:plant_g} describes an asymptotically stable physical system, then:
\color{black}

\begin{equation}\label{eq:stability}
	\lim_{t \longrightarrow \infty} X(t)= X_d(t)
\end{equation}

The demonstration of this property is realized for the studied physical system in Section \ref{sec:asymptoticStab}.
\subsection{Algorithm}\label{sec:algorithm}
The algorithm used to implement the controller given by Eq.s \eqref{eq:ES} and \eqref{eq:ud}, is composed of two main tasks. The first task consists in obtaining the set-point, i.e. integrating the ES of Eq. (\ref{eq:ES}) fed with the measured external forces using a numerical scheme. At each step $t_{n-1}$, the numerical scheme allows to predict the vector state $X_d$ at the next time step $t_{n}$. The second task follows the first one, i.e. it is performed at time step $t_n$, and consists of writing the control variable $u_d$ according to Eq. \eqref{eq:ud}.
Below, the two steps are detailed, starting from $t_n$. For the sake of clarity, the variables computed at $t_{n-1}$ corresponding to the prediction at $t_n$, will be noted $\bullet^{n-1}_n$.
\begin{enumerate}
	\item From Eq. \eqref{eq:plant_g}, taking $\dot X_{d}\equiv\dot X_{d,n}^n$, $X_d\equiv X_{d,n}^{n-1}$ and $f_{ext}^m\equiv f_{ext}^m(t_n)$, we directly compute $\dot X_{d,n}^n$. Hence, we can get $u_d(t_n)$ by Eq. \eqref{eq:ud}.
	
	\item Solve Eq. (\ref{eq:ES}) using a numerical scheme, based on the state vector determined at the preceding time step $X_{d,n}^{n-1}$, and the measured external forces $f_{ext}^m(t_n)$. This operation gives the state vector $X_{d,n}^{n+1}$.\\
\end{enumerate}

Observe that, if functions $\varphi$ and $F$ are both linear, a Laplace transform allows to get $\tilde{X}_d(s)$ as function of $\tilde{f}_{ext}^{ES}(s)=\tilde{f}_{ext}^m(s)$ where $s$ is the frequency domain complex variable (set $\tilde{\bullet}$ denotes the Laplace transform of variable $\bullet$). Therefore, the controller $\tilde{u}_d(s)$ can be directly related to $\hat{f}_{ext}^m(s)$ by a corrector transfer function $H(s)$, and the two steps described above merge into a single one, which is the digital synthesis of $\tilde{u}_d(s)=H(s)\tilde{f}_{ext}^m(s)$. This digital synthesis can be performed by any recursive scheme such as the Infinite Impulse Response (IIR) \cite{goodwin2001control}. Nevertheless, if either $\varphi$ or $F$ are nonlinear, the convolution operators cannot be employed, and the two-steps described above, involving a real-time integration of the ES, must be retained. For this reason, such model-inversion control strategy has been labelled as Real-Time-Integration (RTI) scheme \cite{DeBono2024,MORELL2024118437,morell2025experimental}, while the choice of the specific integration scheme should be made based upon the experimental hardware at disposal.
\begin{figure}
	\centering
	\includegraphics[width=\linewidth]{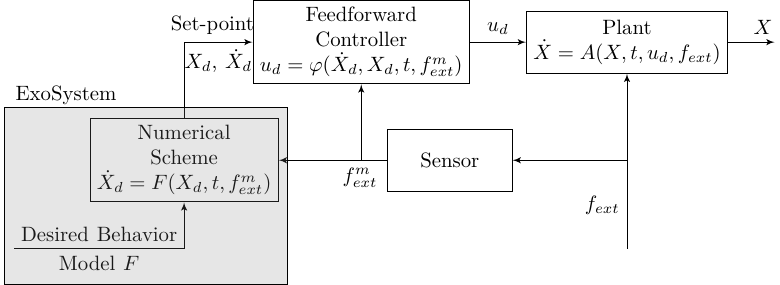}
	\caption{Block diagram of the open-loop control}
	\label{fig:block_diagram}
\end{figure}

\section{Targeting a specific equilibrium point}\label{sec:targeting a specific equilibrium point}
The main objective of this study is to enable the RTI strategy described above, to target specific stable equilibrium points if more than one coexist at a single frequency.
Nonlinear mechanical systems can present complicated basin of attraction which depend on initial conditions \cite{TureSavadkoohi2011}. As a result, depending on initial conditions, the forced nonlinear system can be attracted by different equilibria which affects dramatically the performances of the system. However, on the one hand, to mechanically impose initial conditions on an ER is not a trivial task. On the other hand, classical control strategies do not contemplate the enforcing of initial conditions and the synthetic selection of one specific set-point when the system allows multiple stable solutions. The control strategy presented in Sections \ref{sec:conditions} and \ref{sec:general formulation} allows to tackle this issue by forcing the ES to reach a specific equilibrium point, and is the main novelty of this contribution. Such achievement provides a significant breakthrough in the recent development of non-linear (generalized) impedance control techniques \cite{guo2020improving,DeBono2024,MORELL2024118437,morell2025experimental}, since the specific set-points introduced by synthetic nonlinearities have been, so far, only achievable under unrealistic external noise excitations (swept sines).\\
Let us identify a target equilibrium point $X_d^*$ of the desired dynamics $F$ of the ES. Its basin of attraction is denoted $\mathcal{B}_F(X_d^*)$. In classical nonlinear systems where initial conditions are physically accessible, they can be properly set based on its basin of attraction $B_F(X_d^*)$, in order to realize the target equilibrium point \cite{bellet2010experimental}. For example, supposing to have multiple stable solutions and to target the highest energy equilibrium point, non-zero initial conditions can be forced by pre-constraining the system in order for it to settle on the desired high energy equilibrium point. During an active control process, accessing and imposing the initial conditions is a challenging task, in addition to the basin of attraction that changes as a function of the forced excitation amplitude. However, the ES concept allows to dissociate the external force considered in the ES ($f_{ext}^{ES}$) from $f_{ext}^m$, and consider an $f_{ext}^{ES}=f_{ext}^m+f_{ext}^a$ in the ES, with $f_{ext}^a$ an additional purely synthetic external force, which has the role to lead the ES to a fixed point lying in the basin of attraction $\mathcal{B}_F(X_d^*)$. The easiest way to target a stable fixed point is to drive the behavior of the system to the targeted fixed point. The additional purely synthetic external force $f_{ext}^a$ should have transient character, such that to quickly extinguish after the desired equilibrium point has been reached by the plant. Notice that the extensions of the basin of attraction of each stable solution change with the amplitude of the actual external excitation $f_{ext}$, and the system is not autonomous. Nevertheless, if the amplitude of $f_{ext}^a$ is properly devised based upon the amplitude of $f_{ext}^m$, the basin of attraction of the desired equilibrium can extend such that to fill the entire space of initial conditions physically realisable, and the achievement of the desired set-point is assured. The strategy is described in better details below.\\


By contemplating $f_{ext}^{ES}\neq f_{ext}^m$, the modified controlled system of Eq. \eqref{eq:global}, rewrites:
\begin{subequations}
	\begin{equation}
		\dot X(t)=A(X(t), t, f_{ext}(t), u_d(t))
	\end{equation}
	\begin{equation}\label{eq:ES_fext}
		\dot X_d (t)= F(X_d(t),t,f_{ext}^m(t)+f_{ext}^a(t))
	\end{equation}
	\begin{equation}
		u_d(t)=\varphi(\dot X_d(t),X_d(t),t,f_{ext}^m(t))
	\end{equation}
\end{subequations}
Solving the ES Eq. (\ref{eq:ES_fext}) yields the vector states $X_d^f$ as a solution function of the variables $t\in[0,T]$ and $f_{ext}^{ES}=f_{ext}^m+f_{ext}^a$. The additional transient $f_{ext}^a$ creates new attracting equilibrium points that correspond to solutions $X_d$ for higher or lower excitation amplitudes. Supposing an additional synthetic force such that $\lim\limits_{t \to +\infty} f_{ext}^a(t)=0$, once the values $X_d$ reach the basin of attraction $\mathcal{B}_F(X_d^*)$ of the targeted equilibrium point $X_d^*$, the equilibrium point adopted by the plant will stay on the desired trajectory, i.e.:
\begin{equation}
	\lim\limits_{t \to +\infty} X_d(t)=X_d^*.
\end{equation}
In particular, the flow of the modified system (with $f_{ext}^{ES}\neq f_{ext}^m$) is equal to the flow of the unmodified dynamical system when $f_{ext}^a$ extinguish. If the flow intersects a stable attracting fixed point of the unmodified system, the system is attracted to it. To efficiently use this method, $f_{ext}^a$ may be conceived as an offset smoothly decreasing to 0 with $t$ to ensure that the flow quickly intersects with the desired equilibrium point.

\section{The case of the electroacoustic resonator}\label{sec:the case of ER}
\subsection{The plant and the controller}
After having outlined the mathematical framework of the control strategy to achieve a desired equilibrium point of a non-linear impedance operator, we consider the specific application of a digitally-programmable ER under low noise excitation amplitudes. At low external sound pressure excitations, the loudspeaker natural dynamics is linear. Moreover, by considering sufficiently low operating frequencies, we can consider the loudspeaker's response as dominated by its first mode (also called piston-mode). Hence, the loudspeaker mechanical dynamics can be described by the one-degree-of-freedom ($n=1$) mathematical model \cite{beranek2012acoustics}:
\begin{equation}\label{eq:LS}
	M_0 \ddot x(t) + R_0 \dot x(t) + K_0 x(t) = f_{ext}(t) - Blu_d(t),
\end{equation}
where the variable $x$ is the modal coordinate of the loudspeaker's first mode. The left-hand-side represents the classical single degree of freedom model with the modal mass $M_0$, damping $R_0$ and stiffness $K_0$. 
The force induced by the sound pressure on the membrane surface is denoted by $f_{ext}(t)=S_d p(t)$ where $p$ is the sound pressure, and $S_d$ is the so-called effective surface area of the loudspeaker's membrane. The pressure at the surface of the membrane of the loudspeaker can be measured using one or more microphones \cite{de2022effect}. The control is applied by the Laplace force $-Blu_d(t)$, where $l$ is the length of a solenoid through which the electric current $u_d$ (the controller) flows, and $B$ is the permanent magnetic field generated by the magnet immersed in the solenoid.\\
The model of the plant satisfies the conditions of the implicit function theorem detailed in Section \ref{sec:conditions}, as the control variable is linear with respect to the model of the plant, see Eq. \eqref{eq:LS}. Hence, from Eq. \eqref{eq:ud}, we can easily write the feed-forward controller expression:
\begin{equation}\label{eq:Ud}
	u_d(t)=\dfrac{1}{Bl}(-M_0 \ddot x_d(t) - R_0 \dot x_d(t) -K_0 x_d(t) +f_{ext}^m(t))
\end{equation}
where $x_d\in\mathbb{R}$ is the desired modal coordinate determined by the ES. Here, the practical invertibility condition writes $Bl\ne 0$, which is naturally fulfilled. Observe that the ES is assumed to have the same number of degrees of freedom $n=1$ as the plant. Now, let us particularize Eq. \eqref{eq:ES} for an ES target dynamics of the type:
\begin{equation}\label{eq:ES_ER}
	M_d \ddot x_d(t) + R_d \dot x_d(t) + F_{NL}(x_d,t)=f_{ext}^m(t),
\end{equation}
where $M_d$ is the desired modal mass, and $R_d$ is the desired modal damping, while $F_{NL}:\mathbb{R}\times[0,T] \longrightarrow \mathbb{R}$ is a nonlinear restoring force function. 
Clearly, the choice of the desired modal parameters and of $F_{NL}$ should respect the classical constraint of causality and stability (bounded response). 


\subsection{Asymptotic stability}\label{sec:asymptoticStab}
In this section, we verify the asymptotic stability condition Eq. (\ref{eq:stability}) for the above mentioned control scheme. 
Assuming that the loudspeaker's model is known (no uncertainties) and that pressure measurements are perfectly accurate $f_{ext}^m=f_{ext}$, and replacing $u_d$ in Eq. \eqref{eq:LS} by Eq. \eqref{eq:Ud}, we have $\forall t \in[0,T]$:
\begin{equation}
	M_0 \ddot x(t) + R_0 \dot x(t) + K_0 x(t) = M_0 \ddot x_d(t) + R_0 \dot x_d(t) + K_0 x_d(t).
\end{equation}
Let us set $y=x-x_d$, it leads to a homogeneous second order ordinary differential equation:
\begin{equation}\label{eq:y}
	M_0 \ddot y(t) + R_0 \dot y(t) + K_0 y(t)=0.
\end{equation}
Assuming two complex conjugate roots $r_1$ and $r_2$ for the characteristic equation of Eq. \eqref{eq:y}, the solution writes:
\begin{equation}
	\forall t \in [0,T] \quad y(t)=\e^{\Re(r_1) t}(A\cos (\Im (r_1) t) + B \sin (\Im (r_1) t)),
\end{equation}
where $\Re(z)$ and $\Im(z)$ stand for the real and imaginary parts of a complex number $z\in\mathbb{C}$, respectively. Since the loudspeaker is positively damped as a physical system, it gives a negative coefficient $\Re (r_1)=-R_0/(2M_0)$. In addition, utilizing the triangle inequality, one may observe the behavior of the solution when $t \longrightarrow +\infty$:
\begin{subequations}
	\begin{equation}\label{eq:y asympt stability}
		\lim_{t \longrightarrow +\infty} \e^{\Re (r_1)t}=0
	\end{equation}
	\begin{equation}\label{eq:cos_sin_bounded}
		\forall t\in [0,T] \cup \{+\infty \}\quad |A\cos (\Im (r_1) t) + B \sin (\Im (r_1) t)|\leq|A|+|B|.
	\end{equation}
\end{subequations}
Eq \eqref{eq:cos_sin_bounded} shows that $y(t)$ is bounded, which means that the product limit is 0. It yields:
\begin{equation}
	\lim_{t \longrightarrow +\infty} y(t)=0 \quad \Rightarrow \quad	\lim_{t \longrightarrow +\infty} x(t) = x_d(t).
\end{equation}
So, the designed control satisfies the condition of asymptotic stability and assures that the loudspeaker dynamics tends to the one conceived for the ES.

\section{Numerical simulations of the control}\label{sec:numerical simulations}
This section provides the numerical simulation to numerically validate the proposed strategy for targeting high energy equilibrium points.
\subsection{Validation of the control with $f_{ext}^a=0$, by a comparison to an analytical solution}
First, let us compare the numerical simulations Eq. \eqref{eq:global} to an analytical solution found by using the harmonic balance method (HBM), with a truncation to the first harmonic. For the moment, we assume $f_{ext}^{ES}=f_{ext}^m$, i.e. $f_{ext}^a=0$. At each time step, the ES given by Eq. \eqref{eq:ES_ER} is integrated using a Runge-Kutta numerical scheme, and its solution $x_d$ is injected in the electrical current given by Eq. \eqref{eq:Ud}, as extensively detailed in Section \ref{sec:algorithm} and in \cite{DeBono2024}. Then, the obtained electrical current is employed in the plant Eq. \eqref{eq:LS} which is integrated using a classical numerical scheme, to provide the ER's simulated response $x$.
\begin{table}[ht]
	\centering
	\begin{tabular}{l c c}
		\toprule
		\multirow{1}{*}{Parameter} & Value  & Unit \\ 
		\midrule
		$M_0$     &$3.89\times10^{-4}$ &$\text{kg}$\\
		$R_0$   &$2.63\times10^{-1}$ &$\text{kg}.\text{s}^{-1}$\\
		$K_0$     &$4.34\times10^{3}$ &$\text{kg}.\text{s}^{-2}$\\
		$Bl/S_d$  &$136.7$ &$\text{kg}.\text{m}^{-1}.\text{A}^{-1}.\text{s}^{-2}$\\
		$S_d$ & $1.3\times10^{-3}$ & $\text{m}^2$\\
		\bottomrule
	\end{tabular}
	\vspace{0.5cm}
	\caption{Thiele-Small parameters of the ER.}
	\label{table_ER_0}
\end{table}

\begin{figure}[ht]
	\centering
	\includegraphics[width=0.45\linewidth]{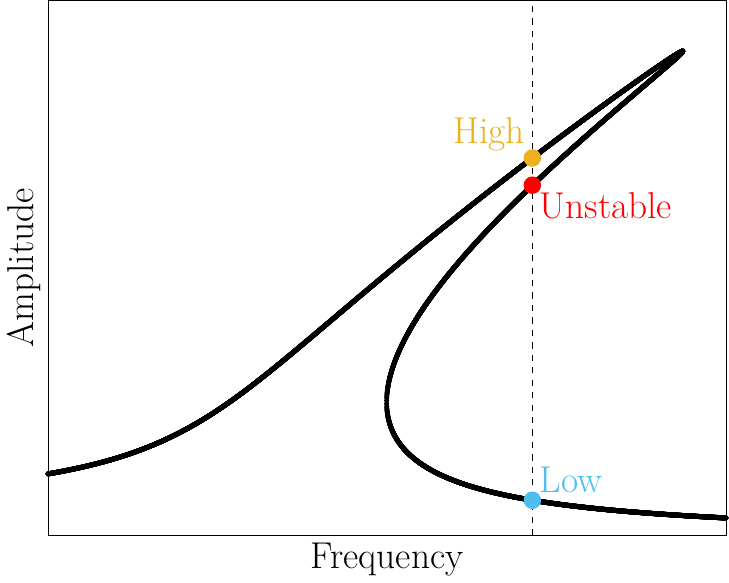}
	\caption{Amplitude response of the Duffing equation featuring a hardening behavior with linear and cubic stiffness.}
	\label{fig:explanation}
\end{figure}

\begin{figure}
	\centering
	\begin{subfigure}[t]{0.49\columnwidth}
		\includegraphics[width=\linewidth]{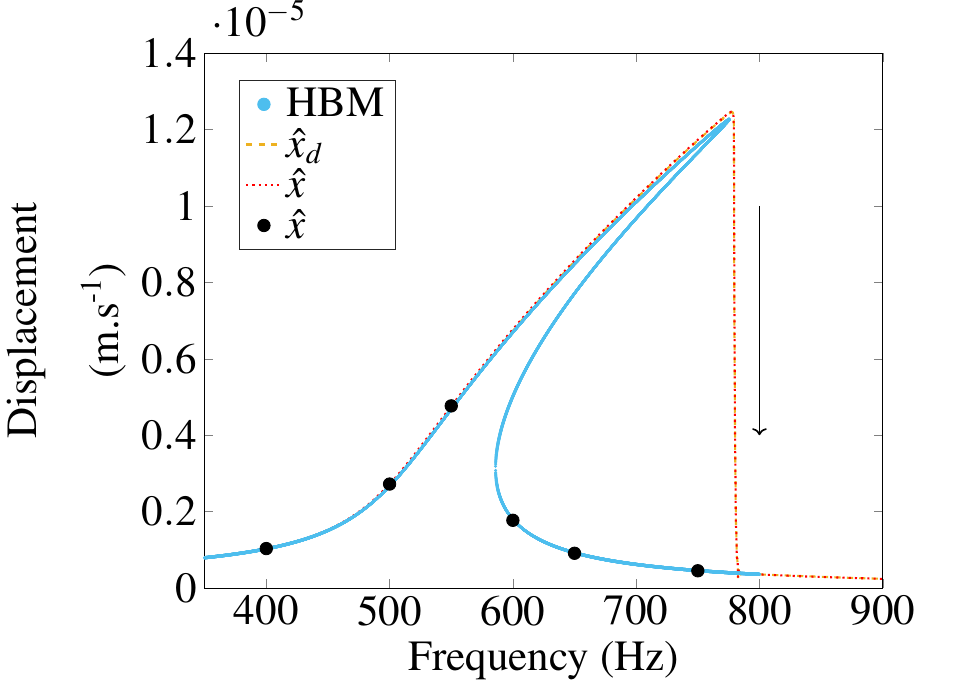}
		\subcaption{}\label{fig:compar_asc_simu}
	\end{subfigure}
	\hfill
	\begin{subfigure}[t]{0.49\columnwidth}
		\includegraphics[width=\linewidth]{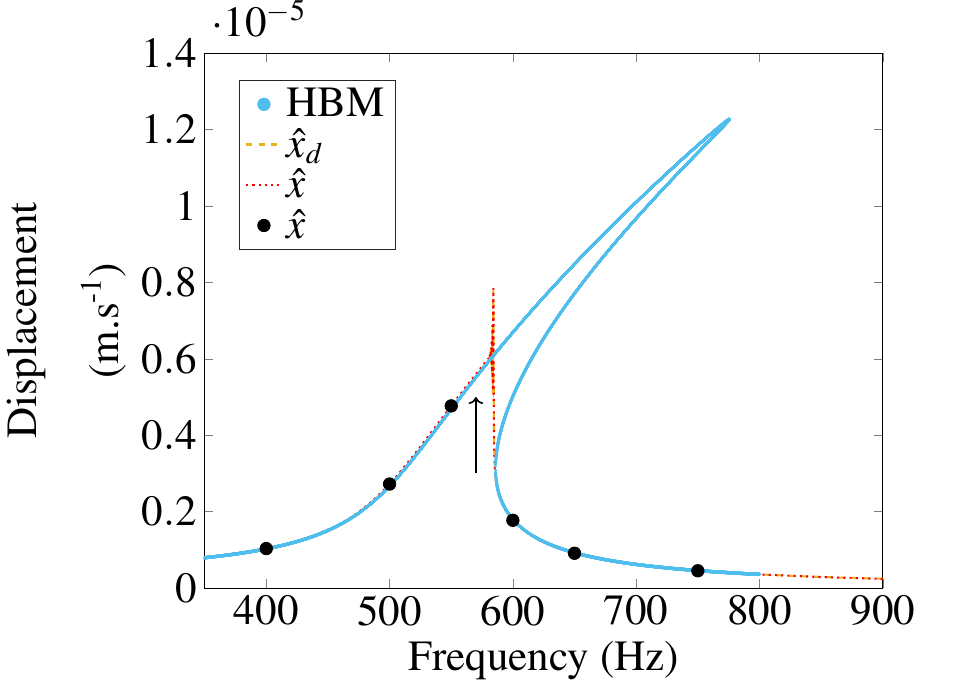}
		\subcaption{}\label{fig:compar_desc_simu}
	\end{subfigure}
	\caption{Comparison between the analytical solution of Eq. \eqref{eq:ES} obtained by Harmonic Balance Method (HBM), and the envelopes $\hat{x}_d$ and $\hat{x}$ of the simulated ES targeted displacement $x_d$ and of the ER response $x$ respectively, in case of (a) increasing or (b) decreasing frequency sweep excitation.}
	\label{fig:simulated frequency responses}
\end{figure}

The Thiele-Small parameters \cite{beranek2012acoustics}, to insert in the plant Eq. \eqref{eq:LS}, ER are given in Table \ref{table_ER_0}. The ES dynamics considered in this paper is a Duffing-like oscillator, with $F_{NL}(x_d,t)=K_d x_d(t) + K_d \beta x_d^3(t)$, but other nonlinear stiffness functions $F_{NL}(x_d,t)$ might be considered, as done in \cite{MORELL2024118437}. Hence, the ES dynamics is described by the following equation:
\begin{equation}\label{eq:ES_exp}
	M_d \ddot x_d(t) + R_d \dot x_d(t) + K_d x_d(t) + K_d \beta x_d^3(t)=f_{ext}^{ES}(t),
\end{equation}
where $M_d$, $R_d$ and $K_d$ denote the desired modal parameters of the ES dynamics. The parameters are given in Eq. \eqref{eq:parameters}, and $\beta=10^{10}\ \text{m}^{-2}$ is the coefficient of the cubic stiffness.
\begin{equation}\label{eq:parameters}
	\begin{aligned}
		&M_d=M_0\\
		&R_d=\dfrac{R_0}{8}\\
		&K_d=K_0.
	\end{aligned}
\end{equation}
We define the parameter $\omega_0=\sqrt{K_d/M_d}$.
The excitation is chosen as a frequency sine sweep lasting 30 seconds, from 350 Hz to 900 Hz (to contain the first mode of the plant), and the excitation amplitude is arbitrarily chosen as $1.5\ \text{Pa}$. Additional simulations are made with a mono-frequency sine signal excitation lasting 5 seconds. The initial conditions for all the simulations are taken as $x(0)=x_d(0)=0$ and $\dot x(0)=\dot x_d(0)=0$.\\
The envelope $\hat{Y}$ of a time signal $Y$ is defined as the magnitude of the Hilbert transform $H$ of the time histories:
\begin{equation}
	\hat{Y}(f(t))=|H(Y)(t)|.
\end{equation}
As the sine sweep associates an excitation frequency at each time $t$, the envelope can be expressed as a function of the excitation frequency denoted $f$: $\hat{Y}(f(t))$.\\
Let us define the high, low and unstable equilibrium points of the response of a Duffing equation featuring a hardening behavior as the defined points in Fig. \ref{fig:explanation}.
The frequency spectra in the stationary regime of $\hat{x}_d$ and $\hat{x}$, corresponding to the amplitudes of the target displacement $x_d$ and of the response of the ER $x$, respectively, are plotted in Fig. \ref{fig:simulated frequency responses}. Both $\hat{x}_d$ and $\hat{x}$ are compared to the HBM solution of Eq. \eqref{eq:ES} for an increasing frequency sweep (Fig. \ref{fig:compar_asc_simu}) and a decreasing frequency sweep (Fig. \ref{fig:compar_desc_simu}). It can be seen that the solution to the ES equation is well solved using the numerical scheme, as it perfectly fits the results obtained from HBM. Both Fig. \ref{fig:compar_asc_simu} and Fig. \ref{fig:compar_desc_simu} show that $\hat{x}_d$ perfectly overlaps with $\hat{x}$, validating the control algorithm in simulations.
Moreover, the ER's stationary response under mono-harmonic excitations (represented by black points) shows that the low energy equilibrium point is the steady-state solution for the considered initial conditions.\\

\begin{figure}[ht]
	\centering
	\includegraphics[width=0.5\linewidth]{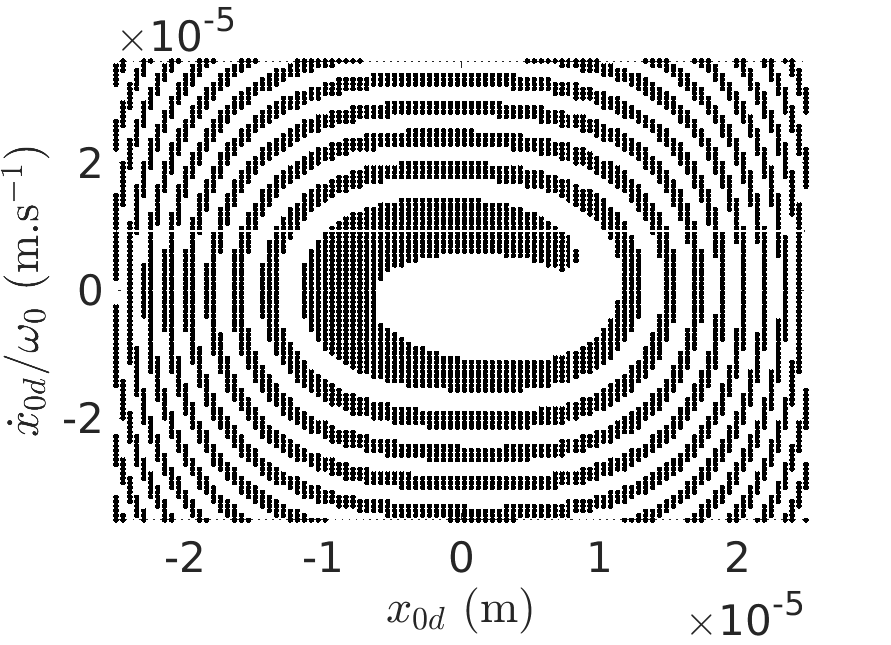}
	\caption{Basins of attraction of the ES Eq. \ref{eq:ES_fext}. The black (respectively white) color zone indicates that the system tends to the high (respectively low) energy equilibrium point.}
	\label{fig:Poincare_ES}
\end{figure}

In Fig. \ref{fig:Poincare_ES}, the basins of attraction of the ES are computed using a Poincare map, depending on the initial conditions $x_{0d}=x_d(0)$ and $\dot x_{0d}=\dot x_d(0)$, for a monoharmonic excitation at $650\ \text{Hz}$ and with the same parameters as the ones of the simulations. The black zone leads to the higher energy equilibrium point, while the white zone leads to the low energy equilibrium point. Figure \ref{fig:Poincare_ES} shows that for initial conditions in the vicinity of the ones chosen in the simulations $(x_0,\dot x_0)=(0,0)$, the system's response tends to the low energy equilibrium point, in agreement with Fig. \ref{fig:simulated frequency responses}.

\subsection{Targeting a specific equilibrium point: simulations}\label{sec:basins}

\begin{figure}
	\centering
	\includegraphics[width=0.7\linewidth]{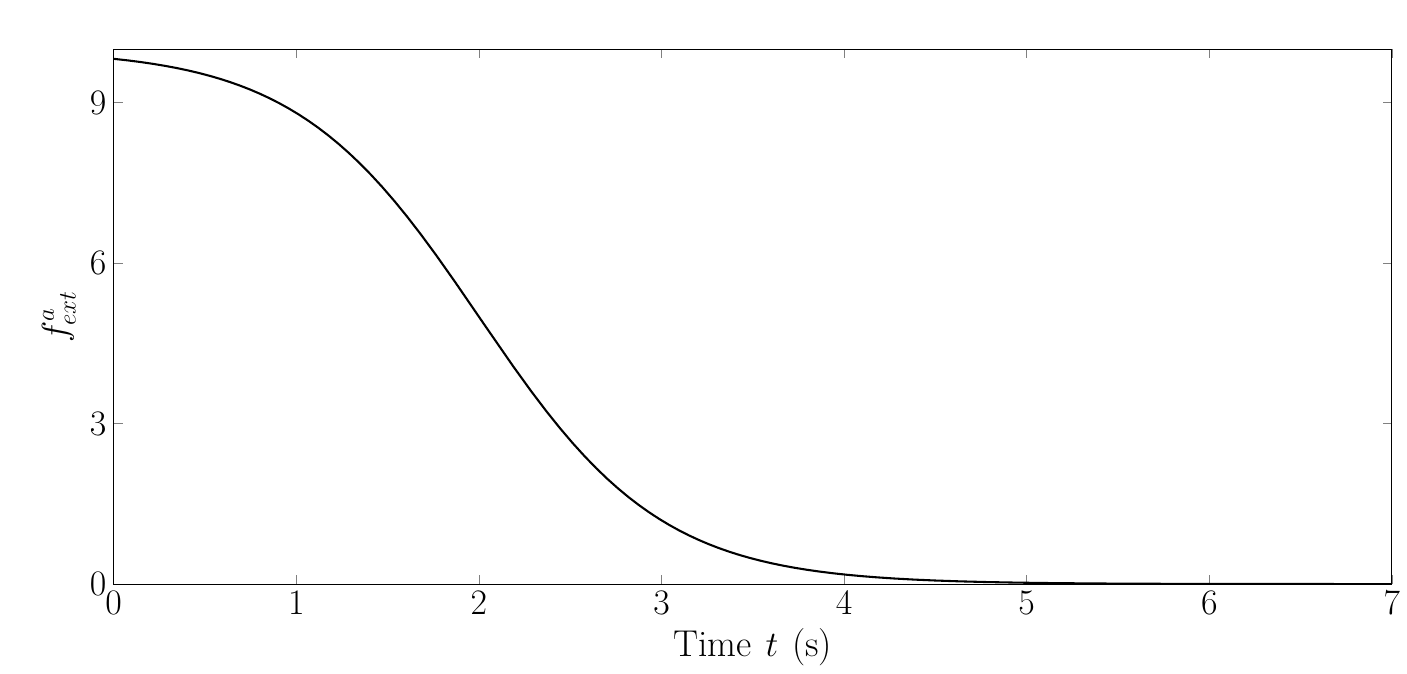}
	\caption{Plot of the additional synthetic transient $f_{ext}^a$.}
	\label{fig:augm}
\end{figure}

Here, we numerically validate the technique described in Section \ref{sec:targeting a specific equilibrium point} to target a specific equilibrium point independently of the initial conditions. 
As presented in Section \ref{sec:targeting a specific equilibrium point}, the additional synthetic transient excitation $f_{ext}^a$ needed to reach the desired equilibrium point, can be conceived as an offset smoothly decreasing to 0 with $t$, as in Eq. \eqref{eq:form_fictive}:
\begin{equation}\label{eq:form_fictive}
	f_{ext}^a=S_d \dfrac{a}{2}\big(1-\tanh(bt-2)\big)
\end{equation}
where $a$ is the amplitude of $f_{ext}^a$ at $t=0$ (the offset). The delay in which the system attains the equilibrium point can be optimized by the choice of $b$ that determines the slope with which $f_{ext}^a$ tends to the asymptotic value 0. For this study, $b$ is set to 1. The additional synthetic transient $f_{ext}^a$ is plotted in Fig. \ref{fig:augm} for $a=10$.\\

\begin{figure}
	\centering
	\begin{subfigure}[t]{0.49\columnwidth}
		\includegraphics[width=\linewidth]{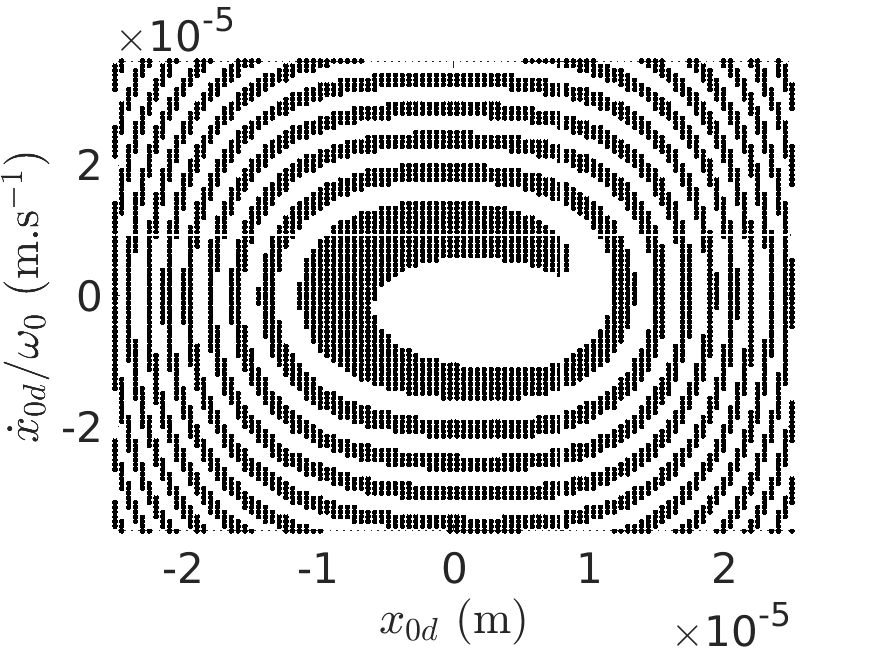}
		\subcaption{$a=2$}\label{fig:BA_ES_p1}
	\end{subfigure}
	\hfill
	\begin{subfigure}[t]{0.49\columnwidth}
		\includegraphics[width=\linewidth]{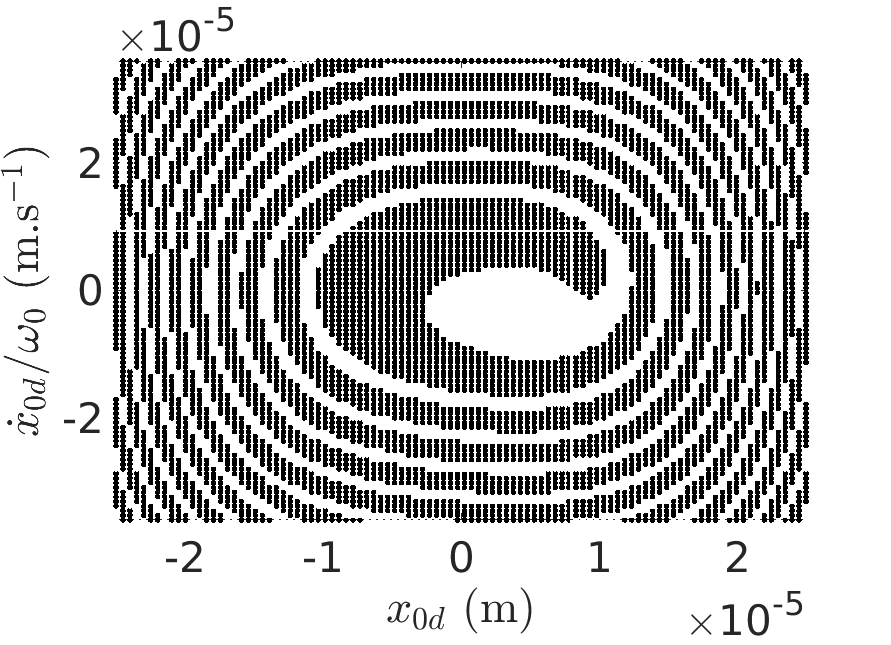}
		\subcaption{$a=10$}\label{fig:BA_ES_p5}
	\end{subfigure}\\
	\begin{subfigure}[t]{0.49\columnwidth}
		\includegraphics[width=\linewidth]{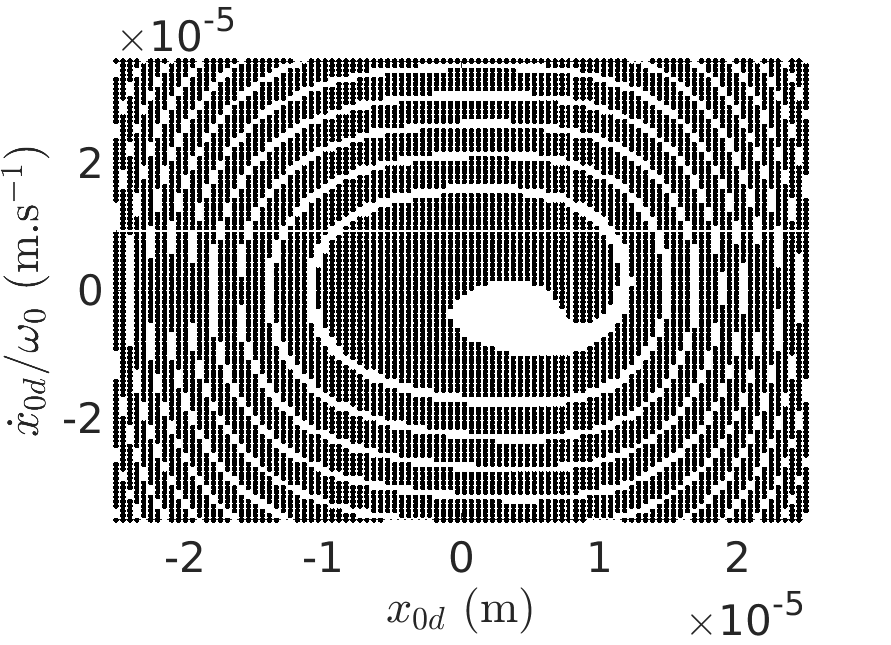}
		\subcaption{$a=12$}\label{fig:BA_ES_p6}
	\end{subfigure}
	\hfill
	\begin{subfigure}[t]{0.49\columnwidth}
		\includegraphics[width=\linewidth]{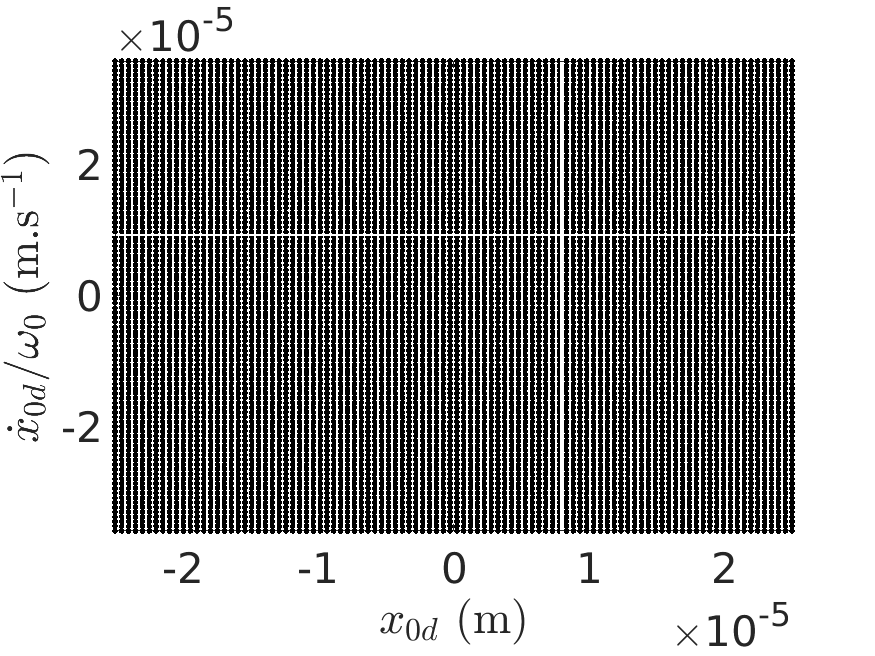}
		\subcaption{$a=14$}\label{fig:BA_ES_p7}
	\end{subfigure}\\
	\caption{Basins of attraction of the ES with the additional synthetic transient $f_{ext}^a$ given by Eq. \eqref{eq:form_fictive}, for different amplitudes $a$. The black (respectively white) colour zone indicates that the system tends to the high (respectively low) energy equilibrium point.}
	\label{fig:Poincare_ES_pulses}
\end{figure}

In Fig. \ref{fig:Poincare_ES_pulses}, the ES basins of attraction for four different values $a$ of $f_{ext}^a$ are plotted, showing that the black zone (corresponding to the initial conditions leading to the high energy equilibrium point) enlarges with $a$. Moreover, a threshold value of $a$ exists such that the entire space of initial conditions contemplated in Fig. \ref{fig:Poincare_ES_pulses}, leads to the desired high energy equilibrium point. Therefore, for a given range of possible initial conditions, the parameter $a$ should be selected higher than the threshold assuring the achievement of the high energy equilibrium point in the entire possible range of initial conditions.\\
Observe that, in case of more than 2 co-existing stable points, the choice of the parameter $a$ is less straightforward, and functions different from the one of Eq. \eqref{eq:form_fictive} might be explored. However, these speculations are left to future investigations which are beyond the scope of the present paper.
\section{The quasi-open field experiment}\label{sec:open field experiment}
The objective is to experimentally validate the nonlinear impedance control strategy introducing an additional synthetic transient $f_{ext}^a$ in the ES, for the selection of the equilibrium point independently of the initial condition of the plant. In this first test, the ER is placed in a quasi-open acoustic field.
\subsection{Experimental set-up}
\begin{figure*}
	\centering
	\begin{subfigure}[t]{0.49\linewidth}
		\includegraphics[width=\linewidth]{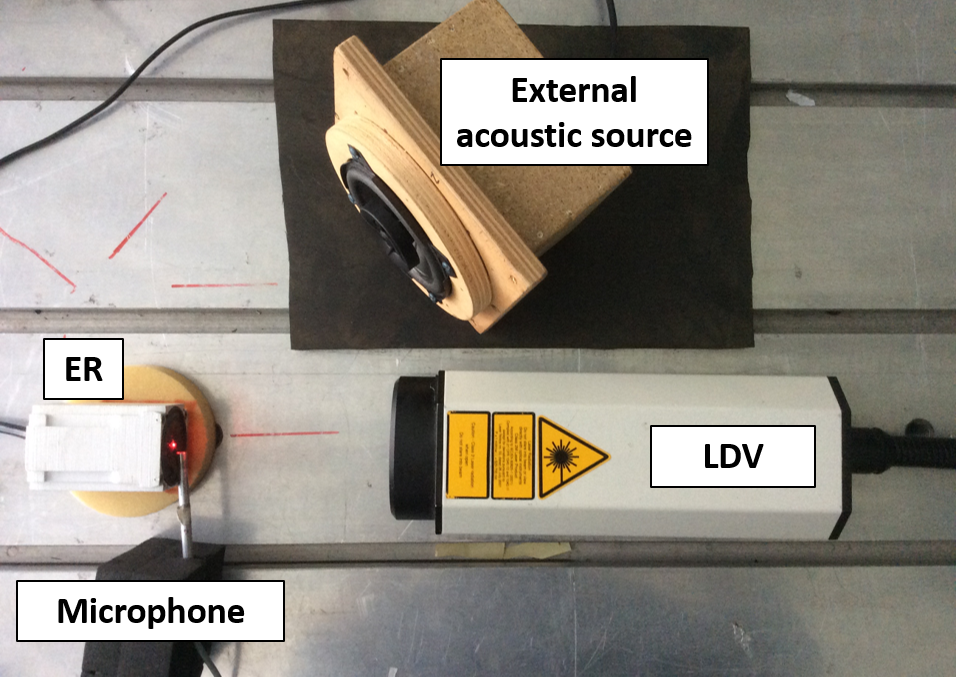}
		\subcaption{}
	\end{subfigure}
	\hfill
	\begin{subfigure}[t]{0.48\linewidth}
		\includegraphics[width=\linewidth]{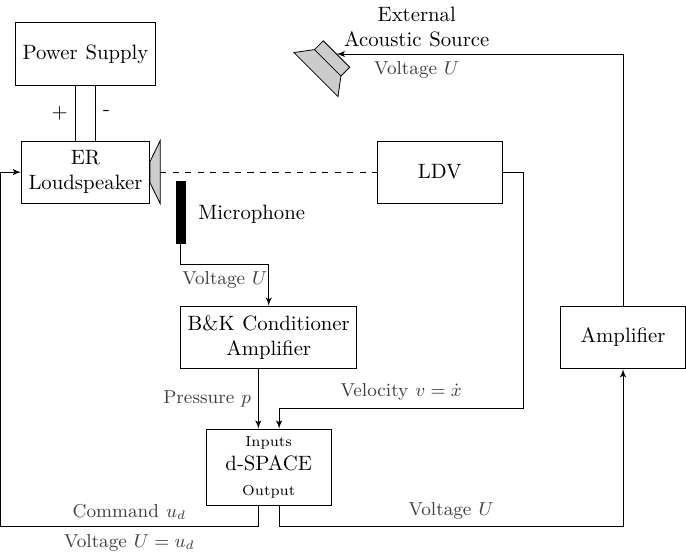}
		\subcaption{}\label{}
	\end{subfigure}
	\caption{(a) Picture and (b) scheme of the open-field experimental set-up.}\label{fig:expOF}
\end{figure*}

The ER is composed of a loudspeaker of membrane effective area $S_d$, collocated to an externally polarized 1/4" Bruel\&Kjaer microphone, and linked to a d-SPACE MicroLabBox DS1202 device for the real-time implementation of the RTI control algorithm. The sampling frequency for both acquisition and control is 50 kHz. The control signal output from d-SPACE provides a voltage command to the loudspeaker proportional to the desired electrical current $u_d$. Indeed, the ER is equipped with a Howland current pump \cite{DeBono2024,DeBono2024j} assuring that the electrical current $u_d$ sent to the loudspeaker's coil is proportional to the voltage command $V_{d}$: \cite{pease2008comprehensive}:
\begin{equation}
	u_d=GV_{\text{d}}
\end{equation}
where the designed Howland current pump sets $G=1\ \text{A.V}^{-1}$.\\
The velocity $v=\dot x$ of the ER's membrane is measured using a Laser Doppler Velocimeter (LDV), while the pressure $p$ on the ER is retrieved by the same microphone employed for the control. The experimental setup is depicted in Fig. \ref{fig:expOF}. The external acoustic source is excited at low amplitudes, such that the ER response would be confined in the linear regime without the controller. Also, the excitation frequencies are sufficiently low to assume a locally-reacting behaviour of the speaker membrane around the piston-mode resonance, as described in Section \ref{sec:the case of ER}. The excitation signals on the external acoustic source can be linearly increasing/decreasing frequency sweeps from 350 Hz to 800 Hz (to check the effect of the initial conditions on the ER synthetic nonlinear dynamics) of 30 seconds, or monoharmonic sines of 5 seconds. In the case of monoharmonic excitations of the external sound source, the ER pressure and velocity are registered for additional 2 seconds, in order to observe the transient regime.\\
The modal parameters of the loudspeaker's first mode and the Laplace force factor $Bl/S_d$ are measured using the method presented in \cite{morell2025experimental}, and are detailed in Table \ref{table_ER_0}.

\subsection{Experimental validation of the control strategy}

\subsubsection{Linear target dynamics}
\begin{figure}
	\centering
	\begin{subfigure}[t]{0.49\columnwidth}
		\includegraphics[width=\linewidth]{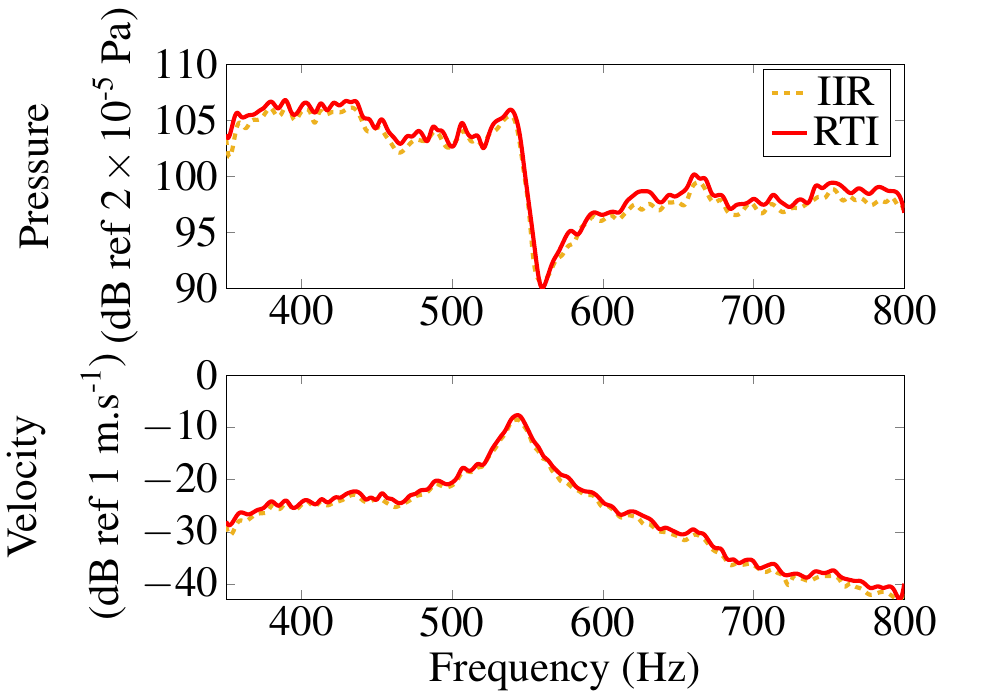}
		\subcaption{}\label{fig:freq}
	\end{subfigure}
	\hfill
	\begin{subfigure}[t]{0.49\columnwidth}
		\includegraphics[width=\linewidth]{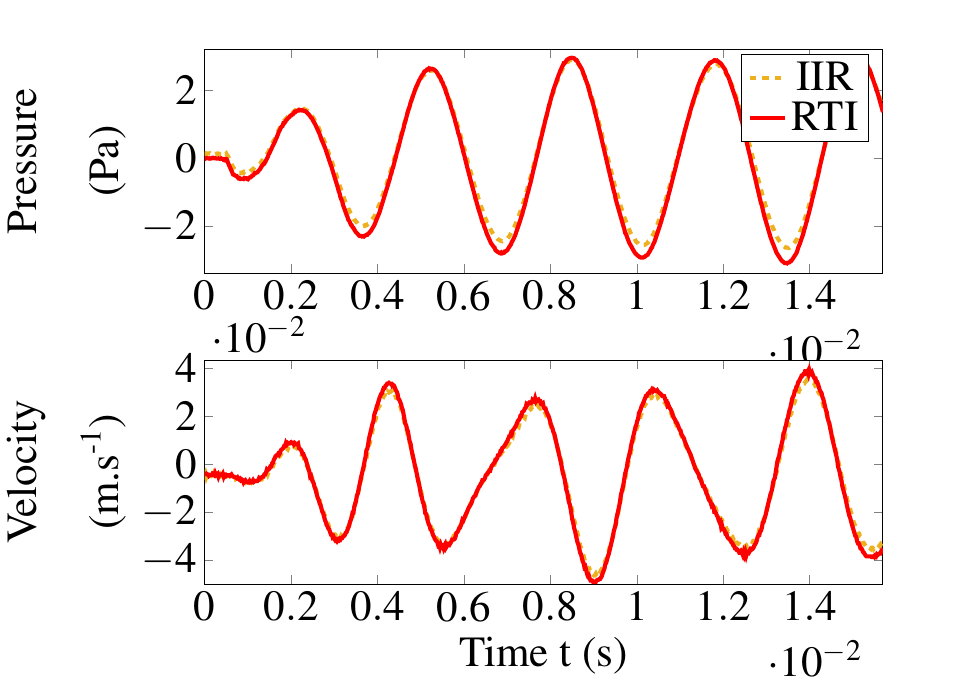}
		\subcaption{}\label{fig:temp}
	\end{subfigure}
	\caption{Comparison of the ER responses in pressure and velocity, controlled by either RTI or IIR algorithms in case of (a) stationary regime, and (b) transient regime.}
	\label{fig:IIR_vs_RTI}
\end{figure}

To validate experimentally the RTI control algorithm, the ER response controlled by the RTI strategy is first compared to the one obtained by classical IIR implementation.
The desired ES dynamics is assumed as a linear SDOF resonator as in Eq. \eqref{eq:linear SDOF target}:
\begin{equation}\label{eq:linear SDOF target}
	M_d \ddot x_d(t) + R_d \dot x_d(t) + K_d x_d(t)=f_{ext}^m(t)
\end{equation}
The transfer function applied through the IIR method is detailed in \cite{DeBono2024}. 
The ER responses are plotted in Fig. \ref{fig:IIR_vs_RTI}, showing very good agreement both for the stationary solutions (Fig. \ref{fig:freq}) within the frequency range of external excitation, and for the transient regime (Fig. \ref{fig:temp}). This first result validates the RTI control for linear target dynamics.\\

\subsubsection{Nonlinear target dynamics}
\begin{figure}
	\centering
	\includegraphics[width=0.5\linewidth]{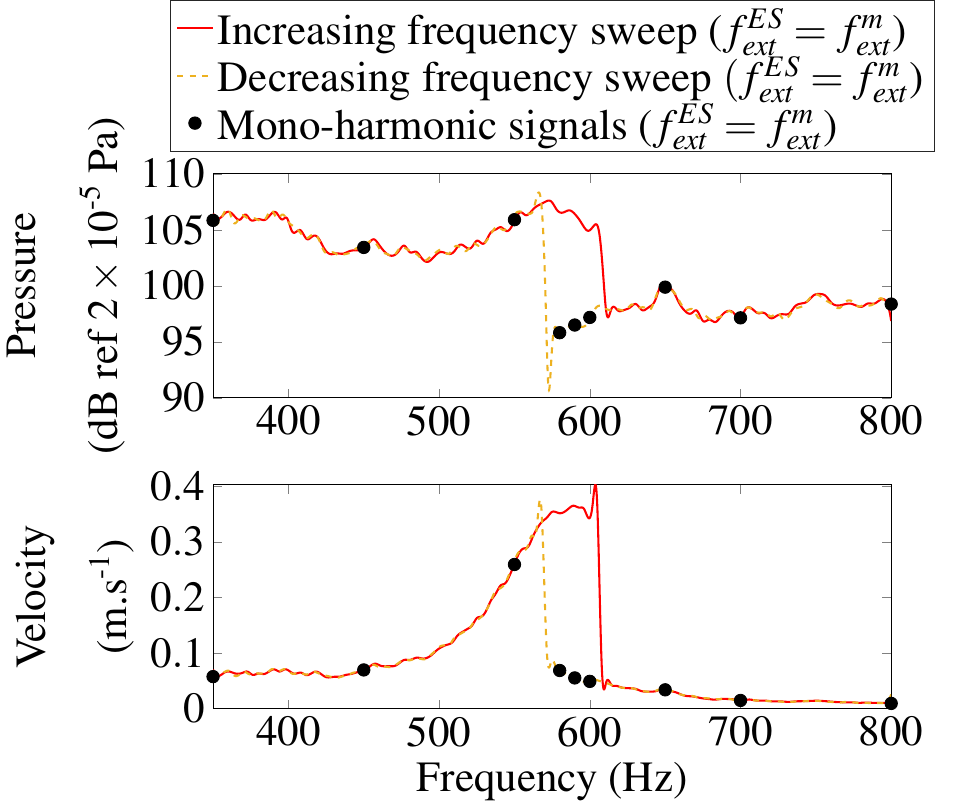}
	\caption{Envelopes of the measured velocity and sound pressure on the ER's membrane, under increasing and decreasing frequency sweep excitations, or under mono-harmonic sine excitations, without employing additional synthetic transients in $f_{ext}^{ES}$.}
	\label{fig:chirp_nonlin}
\end{figure}

In this section, a nonlinear ES is chosen:
\begin{equation}\label{eq:ES_exp2}
	M_d \ddot x_d(t) + R_d \dot x_d(t) + K_d x_d(t) + K_d \beta x_d^3(t)=f_{ext}^{ES}(t).
\end{equation}
For the moment, we assume $f_{ext}^{ES}=f_{ext}^m$ in Eq. \eqref{eq:ES_exp2}.
The desired parameters are listed in Eq. \eqref{eq:parameters}. The amplitude of the parameter $\beta$ can be tuned to ensure that the ER display the nonlinearity under the low external noise excitation levels, while the sign of $\beta$ determines whether the Duffing oscillator has a hardening ($\beta>0$) or softening ($\beta<0$) behaviour. Here, we assume $\beta=10^{10}\ \text{m}^{-2}$, \\
The response of the ER, excited by both increasing and decreasing frequency sweep, in terms of velocity and sound pressure on the ER's membrane, is plotted in Fig. \ref{fig:chirp_nonlin}. In addition, multiple monoharmonic sine signals across the frequency range are also employed as external sound excitations, and the corresponding stationary response envelopes are indicated by dots placed onto the frequency spectra in Fig. \ref{fig:chirp_nonlin}, as done in the simulations of Fig. \ref{fig:simulated frequency responses}. Both pressure and velocity spectra denote a hardening behaviour of the ER, as desired. However, the ER responses to external sine excitations always lie on the lower amplitude branch, i.e. the lower energy equilibrium point, as predicted by the numerical simulations reported in Fig. \ref{fig:simulated frequency responses}. This is so, because in the experimental testing, initial conditions are close to 0 which, as demonstrated by the ES Poincare map of Fig. \ref{fig:Poincare_ES}, reside in the basin of attraction of the lower energy equilibrium point. From the acoustic point of view, the lower branch of the Duffing resonator response corresponds to lower amplitude of the resonance peak and also narrower frequency bandwidth of the ER response, which is undesirable for noise attenuation purposes.\\

\begin{figure}
	\centering
	\includegraphics[width=\linewidth]{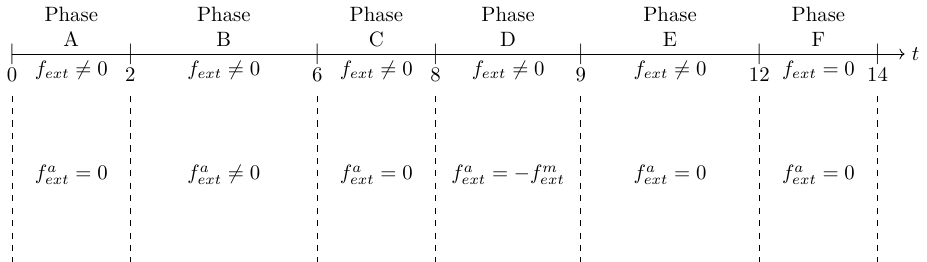}
	\caption{Experimental procedure consisting of a sequence of 6 phases over a 14 seconds measurement.}
	\label{fig:timeline}
\end{figure}

Now, let us consider the possibility of having $f_{ext}^{ES}\neq f_{ext}^m$ and take as external sound excitations, 12 seconds long monoharmonic sine signals, with the ER's sound pressure and velocity measured for 14 seconds, i.e. for 2 additional seconds after the end of external sound excitations. During the 14 seconds-long measurement, different values are assigned to $f_{ext}^a$, in order for the ER to display different behaviours of interest under a sinusoidal external sound excitation. Finally, in the last 2 seconds of the experiment, the external sinusoidal excitation is turned off. Thus, the timeline of the experiment, from 0 to 14 seconds, is divided in different phases, as illustrated in Fig \ref{fig:timeline}, in order to observe the changes in the equilibrium points the ER reaches when $f_{ext}^a$ is varied. During phase A (from $t=0$ to 2 seconds), $f_{ext}^a$ is set to 0, while the ER quickly exhibits a stationary response under the external sine excitation. In phase B (from $t=2$ to 6 seconds), an additional synthetic transient $f_{ext}^a\neq 0$ is taken into account such that the ER can reach the higher energy equilibrium point. Then, during phase C (from $t=6$ to 8 seconds), $f_{ext}^a$ is reset to 0. In phase D (from $t=8$ to 9 seconds) instead, $f_{ext}^a$ is set equal to the opposite of the measured pressure, i.e. $f_{ext}^a=-f_{ext}^m$. This means that $f_{ext}^{ES}=0$, which should soon lead to an $x_d=0$.
Finally, in phase E (from $t=9$ to 12 seconds), the $f_{ext}^a$ is set back to 0, and in phase F (from $t=12$ to 14 seconds) the external sound excitation is turned off.



\begin{figure}[h]
	\centering
	\begin{subfigure}[t]{0.48\linewidth}
		\includegraphics[width=\linewidth]{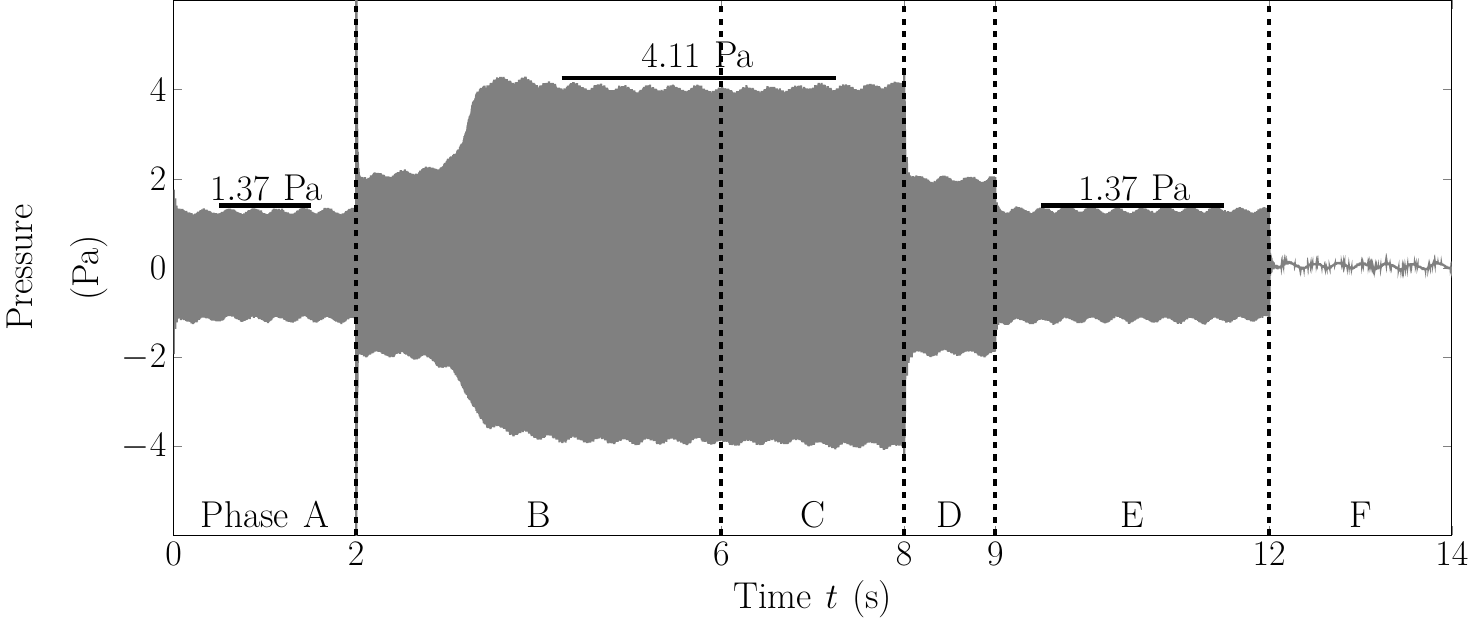}
		\subcaption{}\label{fig:pulse_p_press}
	\end{subfigure}
	\hfill
	\begin{subfigure}[t]{0.48\linewidth}
		\includegraphics[width=\linewidth]{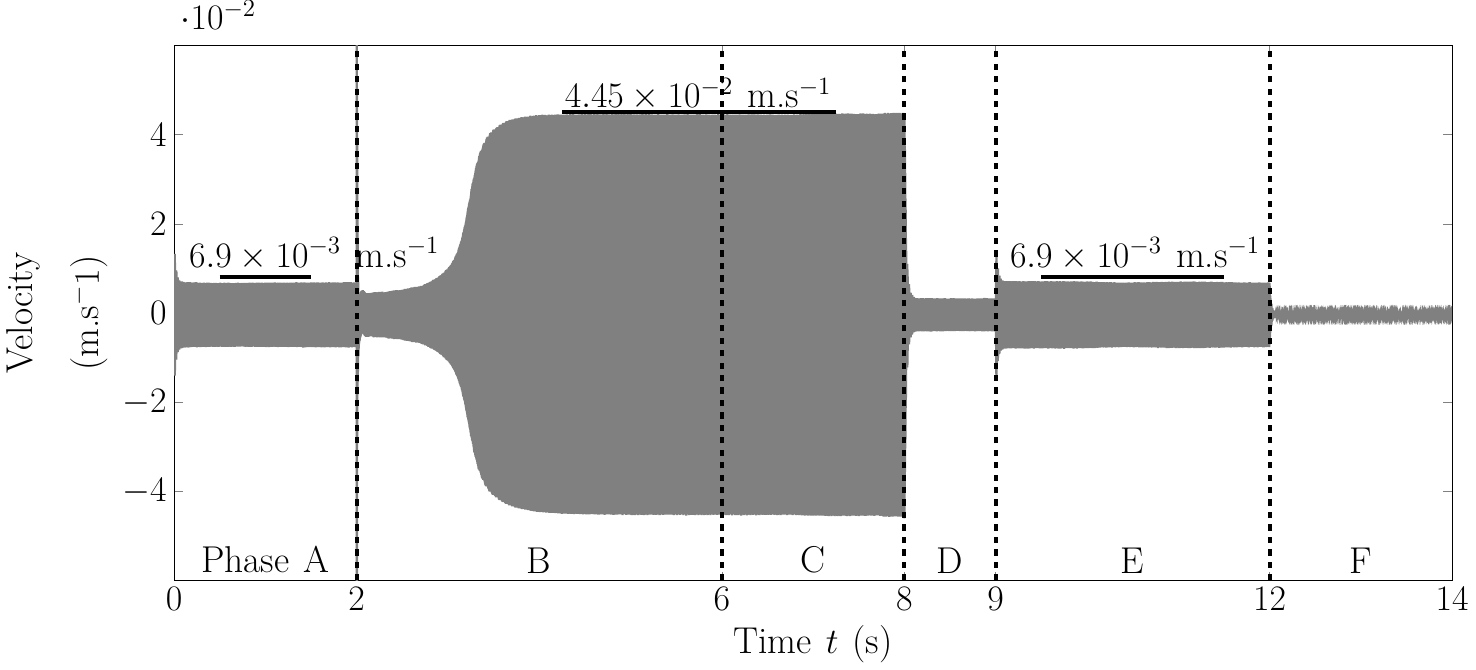}
		\subcaption{}\label{fig:pulse_p_ldv}
	\end{subfigure}
	\caption{Measured sound pressure (a) and velocity (b) on the ER's membrane during the different phases along the timeline of Fig. \ref{fig:timeline}.}
	\label{fig:pulse_p}
\end{figure}

The measured sound pressure and velocity of the ER's membrane during the different phases illustrated in the timeline of Fig. \ref{fig:timeline}, are plotted in Fig. \ref{fig:pulse_p}. The sinusoidal excitation is at 590 Hz. Observe that, during phase A, the lower equilibrium point is attained, as showed previously in Fig. \ref{fig:chirp_nonlin}. Then, in phase B, the higher energy equilibrium point (corresponding to larger oscillation both in pressure and velocity) is reached thanks to an $f_{ext}^a$ given by Eq. \eqref{eq:form_fictive} with $a=200$ Pa. After the $f_{ext}^a$ extinguishes, in phase C, the ER response is stably maintained on the high-energy equilibrium point. Indeed, after the application of the $f_{ext}^a$, the basin of attraction of the high-energy equilibrium point enlarges such that to include the current ES state. So, the ES's response and then the ER's one are attracted and stabilized to the desired equilibrium state, which is therefore endured even after that the additional synthetic transient extinguishes (in phase C).  Then, in phase D we impose $f_{ext}^a=-f_{ext}^m$ which means $f_{ext}^{ES}=0$. Therefore, the ER is forced towards the free response of the ES. This means a minimal velocity despite a significant sound excitation $f_{ext}^m$. Indeed, the actual external excitation is compensated by an $f_{ext}^a=-f_{ext}^m$.
Then, in phase E, $f_{ext}^a$ is brought back to 0, and we retrieve the same ER response as in phase A, as expected. Finally, in phase F, the external sinusoidal excitation is switched off. Notice that, in phase F, $f_{ext}^m$ is almost zero (low background noise) as well as $f_{ext}^a$. This means that $f_{ext}^{ES}=0$ as it was during phase D. Nevertheless, while the measured velocity during phase F falls into background noise, this is not so during phase D, where the velocity oscillations are very low in amplitude but not yet in the background noise levels. This is so because of the inevitable model uncertainties and physiological time delay in the digital control implementation \cite{de2022effect}, for which the model inversion is never perfectly accomplished. This explains the impossibility to reach a rigid ER behaviour ($v=0$) when $f_{ext}^{ES}=0$. 

\begin{figure}
	\centering
	\includegraphics[width=0.5\linewidth]{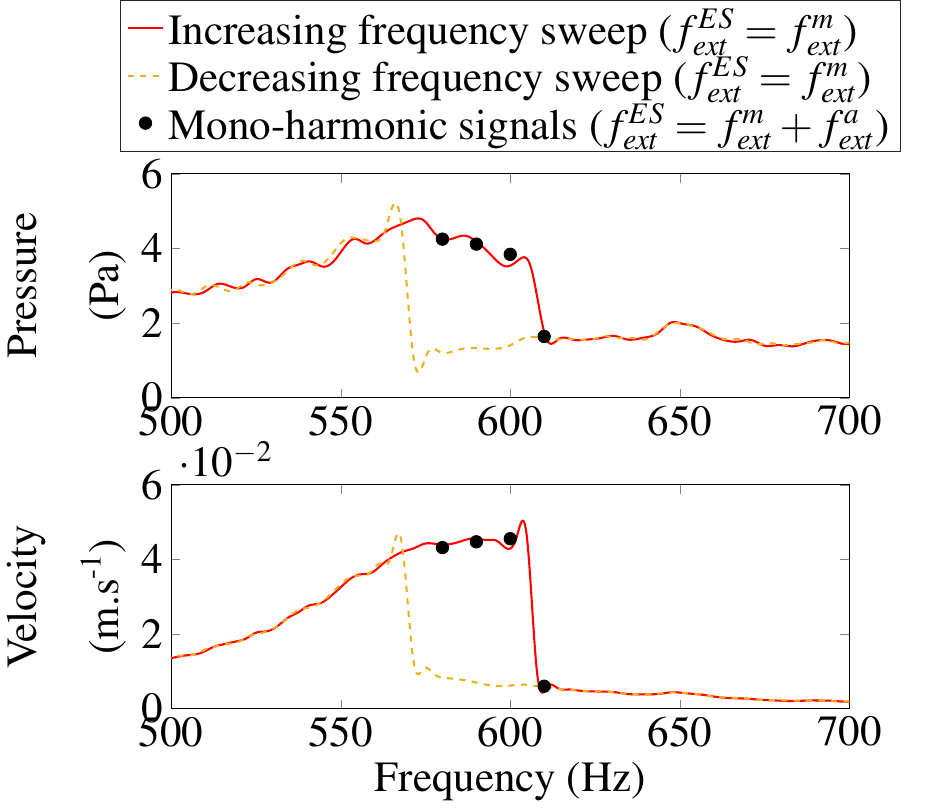}
	\caption{Spectra of sound pressure and velocity amplitudes on the ER's membrane under increasing (solid-line) and decreasing (dashed-line) frequency sweep external excitation, in case of $f_{ext}^a=0$. On the same graphs, pressure and velocity amplitudes of the ER response with $f_{ext}\neq 0$, under sinusoidal excitations at 580 Hz, 590 Hz, 600 Hz, and 610 Hz are displayed as dots.}
	\label{fig:pulse_p_chirp}
\end{figure}

The average values of the envelope of the stationary regime (phase C) for external sinusoidal excitation at four different frequencies (580 Hz, 590 Hz, 600 Hz, and 610 Hz) are reported with dots in Fig. \ref{fig:pulse_p_chirp}. Those points are compared to the spectra obtained with $f_{ext}^a=0$, under increasing and decreasing frequency sweeps. Fig. \ref{fig:pulse_p_chirp} demonstrates that the high-energy equilibrium point has been reached thanks to the additional synthetic transient in the ES dynamics.\\


\section{The coupled-tube experiment}\label{sec:coupled}
In this second experiment, the ER is placed into an enclosed acoustic cavity, corresponding to the experimental testbench reported in \cite{MORELL2024118437}. In \cite{MORELL2024118437}, the RTI algorithm's potential for acoustic mode attenuation, was outlined for different nonlinear synthetic dynamics. Nevertheless, so far, the equilibrium points exhibited by the ER could only be reached by properly adjusting the external sound excitation (increasing or decreasing frequency sweeps, for example), which are unrealistic in actual noise control scenarios. After having assessed the performances of our new control strategy (involving an $f_{ext}^{ES}\neq f_{ext}^{m}$) in quasi-open field, we now test the same concept when the ER is coupled to acoustic cavities.

\subsection{Experimental set-up}
\begin{figure*}[h]
	\centering
	\begin{subfigure}[t]{0.49\linewidth}
		\includegraphics[width=\linewidth]{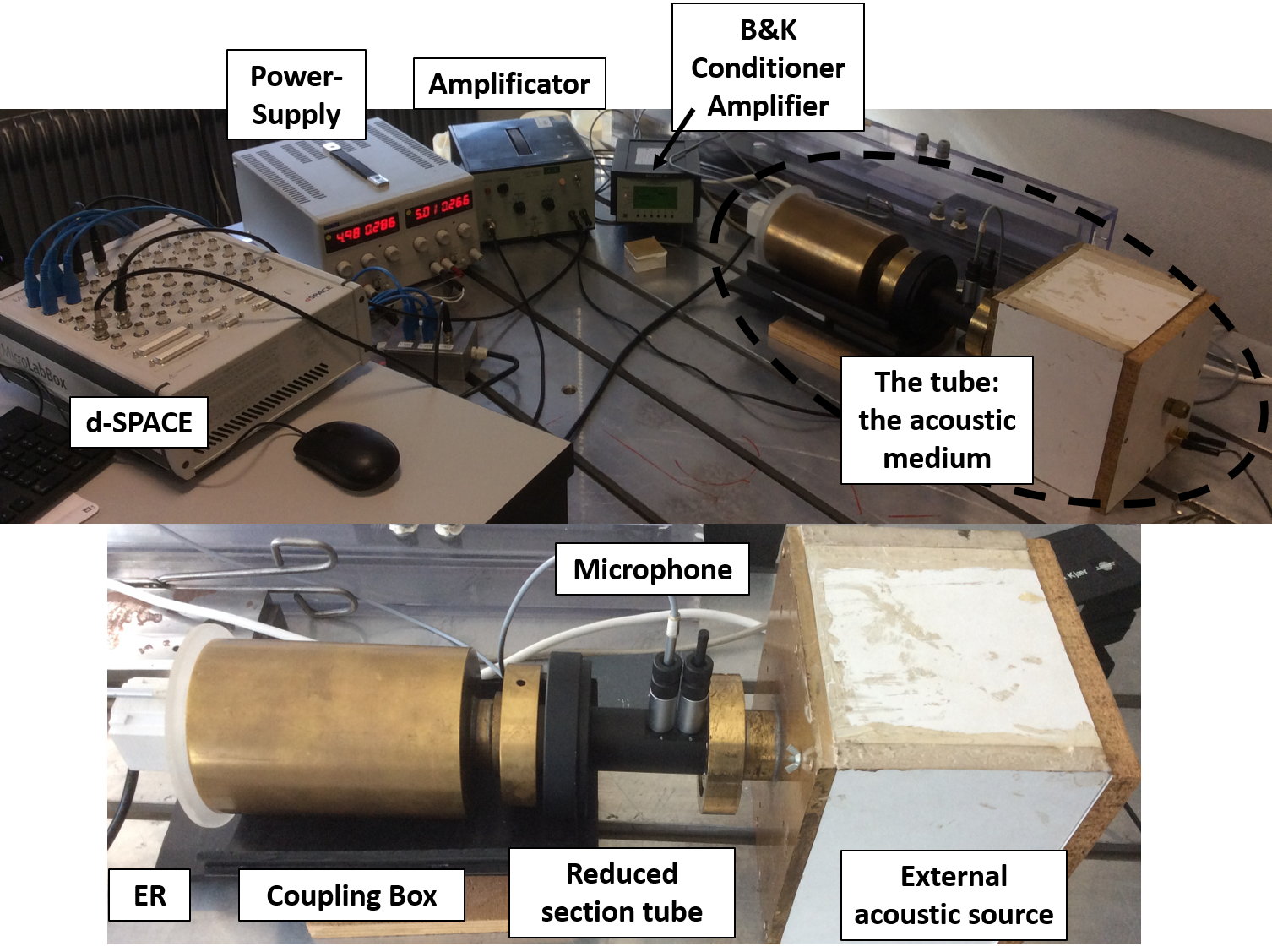}
		\subcaption{}\label{fig:experimental coupled}
	\end{subfigure}
	\hfill
	\begin{subfigure}[t]{0.49\linewidth}
		\includegraphics[width=\linewidth]{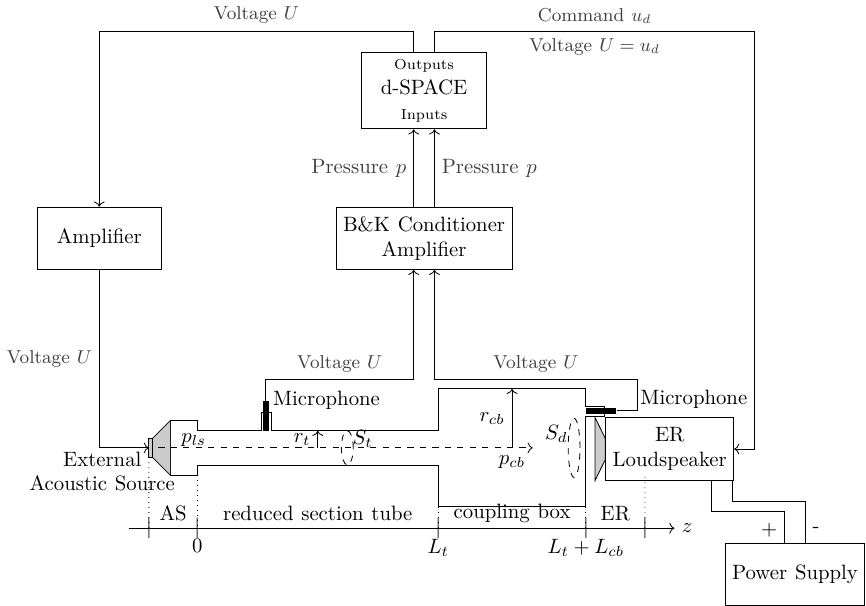}
		\subcaption{}\label{fig:coupled sketch}
	\end{subfigure}
	\caption{(a) Photo and (b) scheme of the experiment with the ER coupled to acoustic cavities.}\label{fig:expCoupled}
\end{figure*}

The ER is placed at the end of a cylindrical cavity with a cross-section $S_{cb}$ (radius $r_{cb}$) and of length $L_{cb}$, which is linked to a cylindrical tube with a narrow cross-section $S_t$ (radius $r_t$) of length $L_t$ at the other end, as pictured in the photos of Fig. \ref{fig:experimental coupled} and in the sketch of Fig. \ref{fig:coupled sketch}. The ER can be replaced with a rigid termination for creating a reference measure. The narrow cross-section tube ($S_t$-tube) is terminated with the acoustic source: an external loudspeaker delivering a sound pressure excitation $p_{ls}$. The $S_t$-tube is considered as the main system whose first acoustic mode must be attenuated by the ER. Below, we report the analytical model employed in \cite{bellet2010experimental,MORELL2024118437,morell2025experimental} to study this specific setup. This model assumes the box of cross-section $S_{cb}$ ($S_{cb}$-cavity) as a weak coupling between the ER and the main system in the frequency range around the first longitudinal mode of the $S_t$-tube. Hence, as the $S_t$-tube is weakly coupled to the ER, we can assume that the variation of the first longitudinal mode shape of the $S_t$-tube is not significantly influenced by the ER dynamics. Hence, a single microphone in the $S_t$-tube is considered as sufficient to identify the modal amplitude \cite{bellet2010experimental}. 

Below, the equations describing such analytical model are provided, which can be useful for the reader to recognize the main parameters involved.


In the following analytical model, the ER dynamics is considered as coincident with the ES, i.e. the model inversion control is supposed to work perfectly. In these assumptions, the experiment can be modeled by the following system of equations:
\begin{subequations}\label{eq:coupled analytical model}
	\begin{equation}\label{eq:ES_coupled}
		m \ddot u(t) + c \dot u(t) + k u(t) +\dfrac{\gamma}{\alpha}(u(t) - \alpha x(t))=-S_t p_{ls}(t)
	\end{equation}
	\begin{equation}\label{eq:acoustic_mode}
		M_d \ddot x_d(t) + R_d \dot x_d(t) + K_d x_d(t) + K_d \beta x_d^3(t)+\gamma (\alpha x(t) - u(t))=0
	\end{equation}
\end{subequations}
where $u$ stands for the modal coordinate of the reduced section tube's first mode, and $m$, $c$ and $k$ represent its modal mass, damping and stiffness. The parameter $\gamma$ is the apparent stiffness of the coupling box that links the ER to the acoustic mode of the narrow tube. Finally, $\alpha=S_d/S_t$. The parameters of the acoustic cavities are estimated according to the procedure described in \cite{morell2025experimental}, and their values are detailed in Table \ref{table:coupled_exp}.
\begin{table}[H]
	\centering
	\begin{tabular}{l c c c}
		\toprule
		&\multirow{1}{*}{Dimension}
		& Value  & Unit \\ 
		\midrule
		\multirow{3}{*}{Tube} & $r_t$ & $1.45\times10^{-2}$ & $\text{m}$ \\
		
		& $S_t$ & $6.61\times10^{-4}$ & $\text{m}^2$ \\
		
		& $L_t$ & $0.204$ & $\text{m}$\\
		
		\cmidrule{2-4}
		\multirow{3}{*}{Coupling box} & $r_{\text{cb}}$ & 0.05 & $\text{m}$ \\
		
		& $L_{\text{cb}}$ & $0.125$ & $\text{m}$ \\
		& $\gamma$ & $125.54$ & $\text{kg.s}^{-2}$ \\
		
		\cmidrule{2-4}
		%
		\multirow{3}{*}{Acoustic mode} & $m$ & $8.22\times10^{-5}$ & $\text{kg}$ \\
		& $c$ & $3.96 \times 10^{-2}$ & $\text{kg.s}^{-1}$ \\
		& $k$ & $1.38\times 10^3$ & $\text{kg.s}^{-2}$ \\
		
		\bottomrule
	\end{tabular}
	\vspace{0.5cm}
	\caption{Dimensions and parameters of the experiment and of its model}
	\label{table:coupled_exp}
\end{table}

Further details about such analytical model can be found in \cite{bellet2010experimental,MORELL2024118437,morell2025experimental}, where the system of Eq. \eqref{eq:coupled analytical model} is integrated. In this paper, the numerical solutions of Eq. \eqref{eq:coupled analytical model} would be redundant, as the objective of this section is to validate the efficiency of the proposed control strategy in an experimental environment (with coupled acoustic cavities \cite{de2022effect}) which might challenge the stability and performance of the control.

\subsection{Targeting nonlinear dynamics}
\begin{figure}[H]
	\centering
	\includegraphics[width=0.5\linewidth]{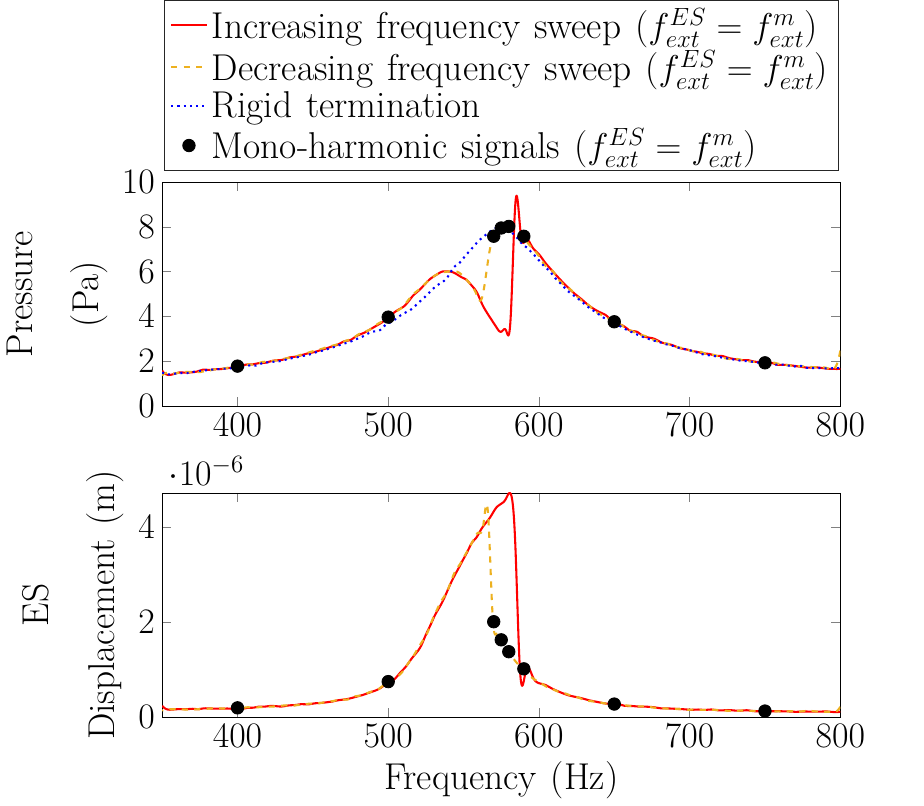}
	\caption{Envelopes of the measured pressure inside the $S_t$-tube, and ES displacement  under increasing and decreasing frequency sweep excitations in case of $f_{ext}^{ES}=f_{ext}^m$, along with the values obtained under mono-harmonic sine excitations.}
	\label{fig:classical_control_coupled}
\end{figure}

The same ES governing Eq. \eqref{eq:ES_exp} is implemented, as for the open field experiment of Section \ref{sec:open field experiment}, whose parameters are given in Eq. \eqref{eq:parameters}. In Fig. \ref{fig:classical_control_coupled}, we report the sound pressure measured inside the $S_t$-tube, and the displacement $x_d$ of the ES estimated by the recursive integration of the ES dynamics (the first step of the control algorithm showed in Section \ref{sec:algorithm}).\\

The envelopes of the measured pressure inside the $S_t$-tube, and the ES displacement are plotted in Fig. \ref{fig:classical_control_coupled}. Frequency sine sweep excitations have been employed with both increasing and decreasing frequencies, and a comparison with a rigid termination measurement is made. Moreover, the stationary responses of the system to single frequency sine signal are plotted for some frequencies. Observe that the ES displacement shows a nonlinear and hardening behaviour. Moreover, the stationary responses to mono-harmonic signals are situated on the lower energy branch. The measured pressure inside the $S_t$-tube shows that adding the ER at the end of the tube reduces the acoustic level in the tube. Observe that the measurements under a decreasing frequency sweep excitation corresponds to lower energy equilibrium points of the ES, and therefore to higher acoustic levels inside the $S_t$-tube. Indeed, when the ER's response is situated in the lower energy equilibrium point, the ER displacement amplitude, speed and acceleration are lower than for the high energy equilibrium points, leading to lower energy transfer from the primary system to the ER. As a result, to activate an efficient transfer energy from the primary system to the ER, and to enlarge the frequency bandwidth of efficient modal attenuation, the ER's response should be situated on the higher energy equilibrium points of the nonlinear resonance. This motivates the introduction of the strategy described above, involving an $f_{ext}^{ES}\neq f_{ext}^m$ to reach the desired equilibrium points.\\

\begin{figure}[H]
	\centering
	\includegraphics[width=0.5\linewidth]{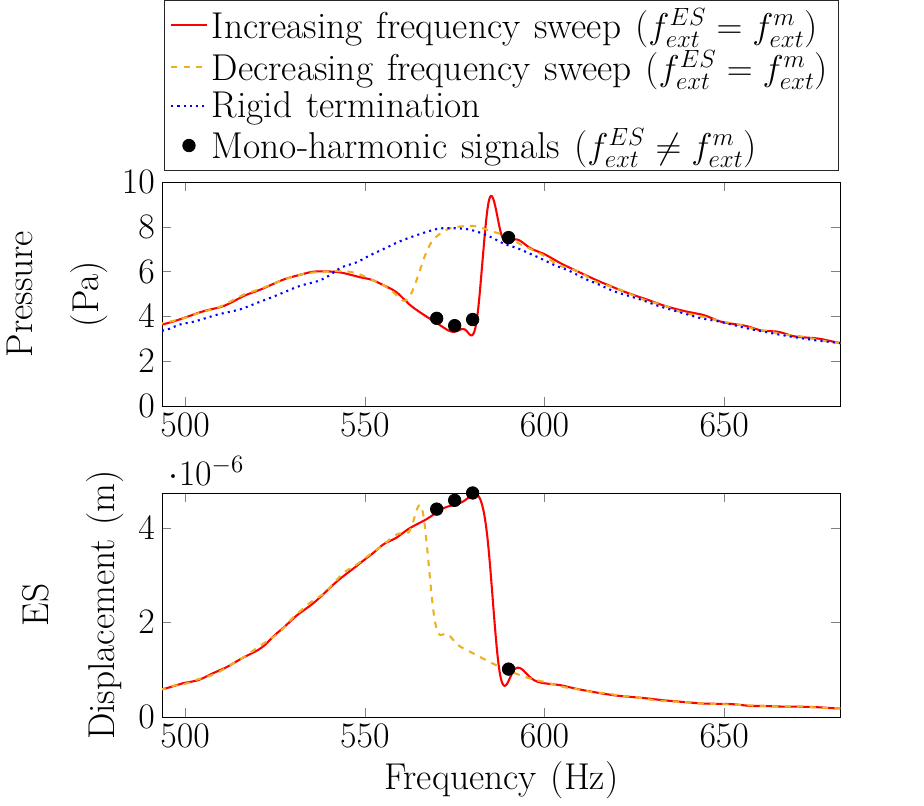}
	\caption{Envelopes of the measured pressure inside the $S_t$-tube, and ES displacement  under increasing and decreasing frequency sweep excitations in case of $f_{ext}^{ES}=f_{ext}^m$, along with the values obtained under mono-harmonic sine excitations during phase C, after the extinguishment of $f_{ext}^a$.}
	\label{fig:modified_control_coupled}
\end{figure}

Now, in order to validate the control strategy involving an $f_{ext}^{ES}\neq f_{ext}^m$, we test the system following the same experimental phases as those reported in the timeline sequence of Fig. \ref{fig:timeline}. The amplitude of the added synthetic parameter is now fixed to $a=200\ \text{Pa}$. The envelopes of the stationary responses to mono-harmonic excitations at 570 Hz, 575 Hz, 580 Hz, and 590 Hz extracted from phases B and C (with the modified control) are plotted in Fig. \ref{fig:modified_control_coupled}. The measured pressure inside the $S_t$-tube and the ES displacement computed in the control loop are plotted. It shows that the additional synthetic transient excitation $f_{ext}^a$ leads the ER's response to attain the higher energy equilibrium point, resulting in better attenuation of the acoustic mode.\\

\begin{figure}
	\centering
	\begin{subfigure}[t]{0.49\linewidth}
		\includegraphics[width=\linewidth]{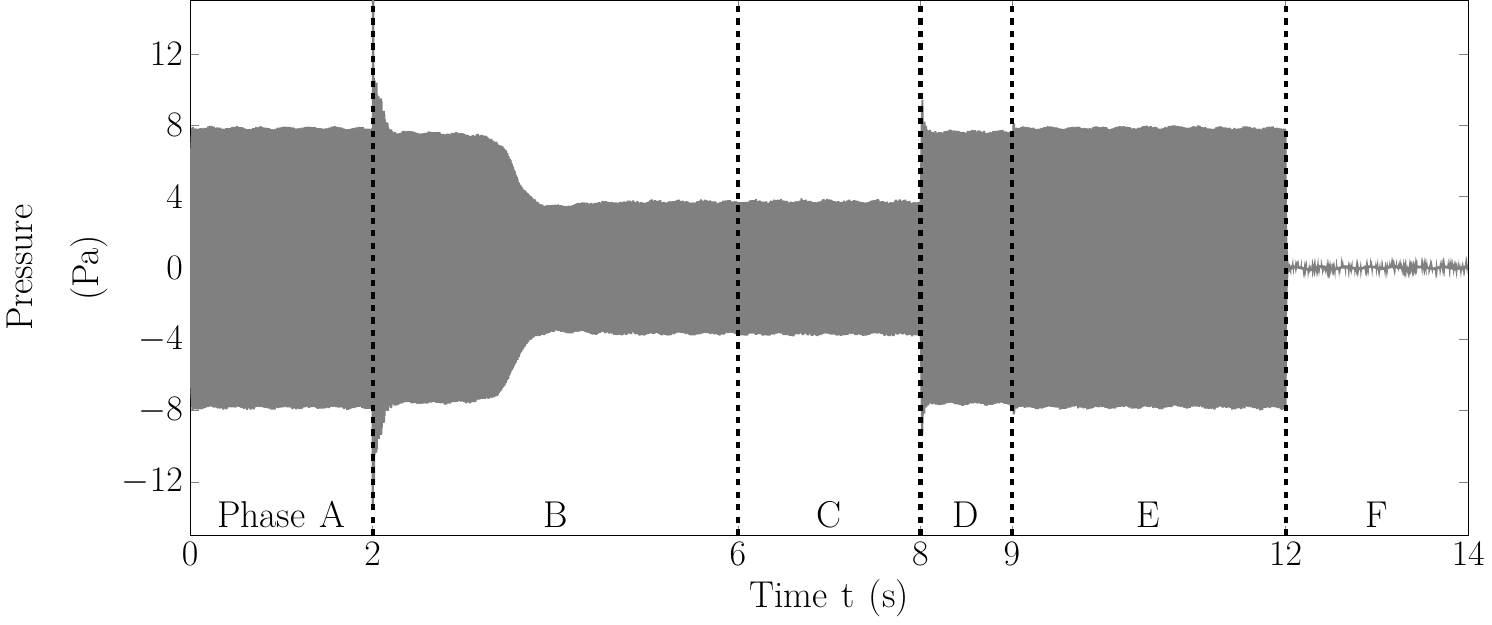}
		\subcaption{}\label{fig:pulse_coupled_p}
	\end{subfigure}
	\hfill
	\begin{subfigure}[t]{0.49\linewidth}
		\includegraphics[width=\linewidth]{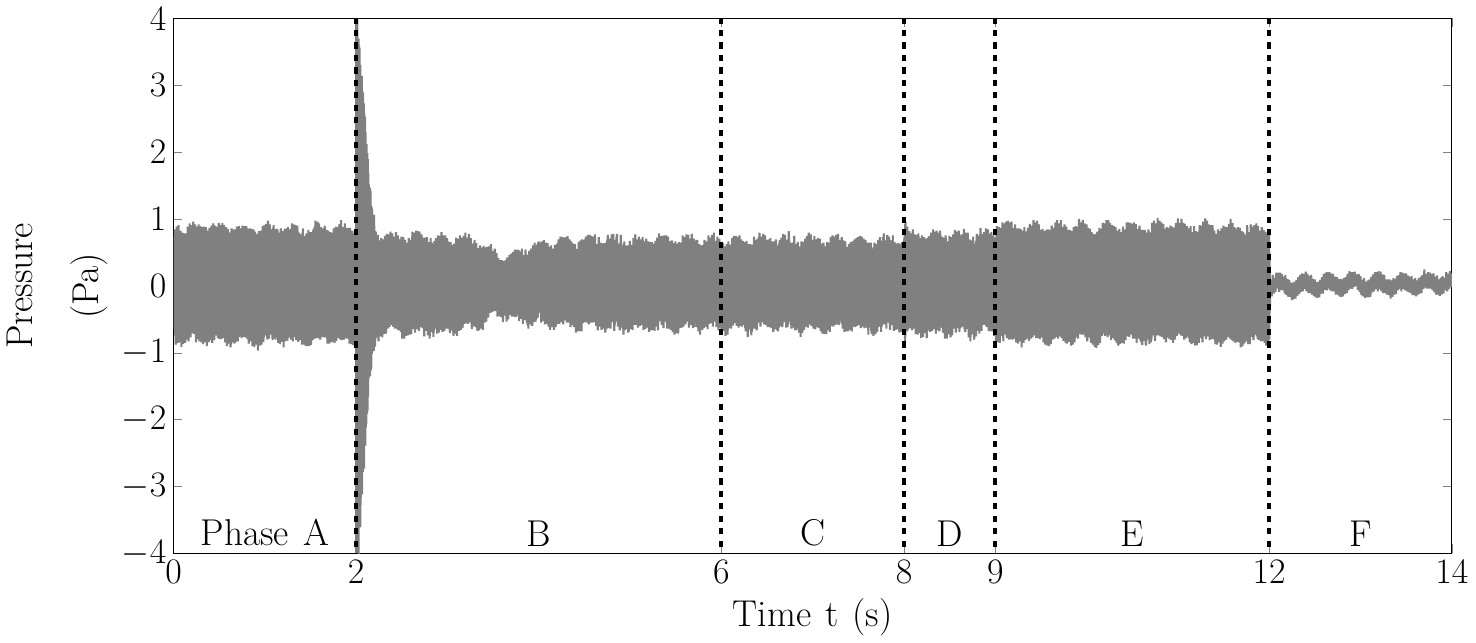}
		\subcaption{}\label{fig:pulse_coupled_pER}
	\end{subfigure}
	\hfill
	\begin{subfigure}[t]{0.49\linewidth}
		\includegraphics[width=\linewidth]{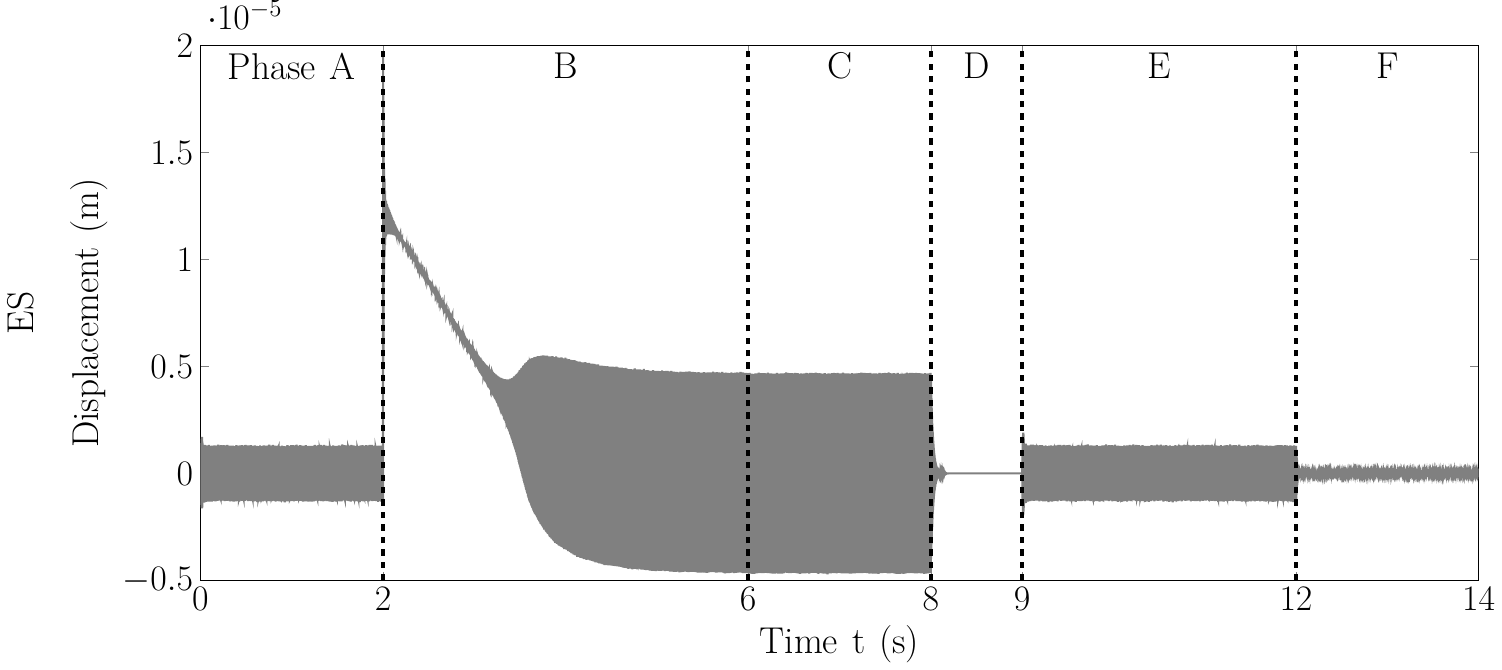}
		\subcaption{}\label{fig:pulse_coupled_displ}
	\end{subfigure}
	\caption{Measured pressures (a) inside the tube with a reduced section and (b) inside the coupling box at the ER's membrane surface for a 580 Hz mono-harmonic signal ; (c) Estimated displacement of the ER's membrane computed in the control loop for a 580 Hz sine excitation signal.}
	\label{fig:pulse_p_coupled}
\end{figure}

The time-histories of the measured pressure in the $S_t$-tube, of the measured pressure inside the $S_{cb}$-cavity, and of the ES displacement, are represented in Fig. \ref{fig:pulse_p_coupled} for a 580 Hz sine excitation. In phases B and C, it can be observed that the high energy equilibrium point in the ER's membrane displacement is attained, demonstrated by lower energy of the acoustic mode's response. The lower energy equilibrium point of the ER, instead, is reached in phase E in accordance with the results of the open-field experiment. Notice that, at the beginning of phase B we can observe an undesirable sudden increase of the ES displacement and of the measured pressures. However, this effect can be significantly reduced by properly adjusting the $a$ parameter in the expression of $f_{ext}^a$ of Eq. \eqref{eq:form_fictive}.

\begin{figure}[H]
	\centering
	\begin{subfigure}[t]{0.49\columnwidth}
		\includegraphics[width=\linewidth]{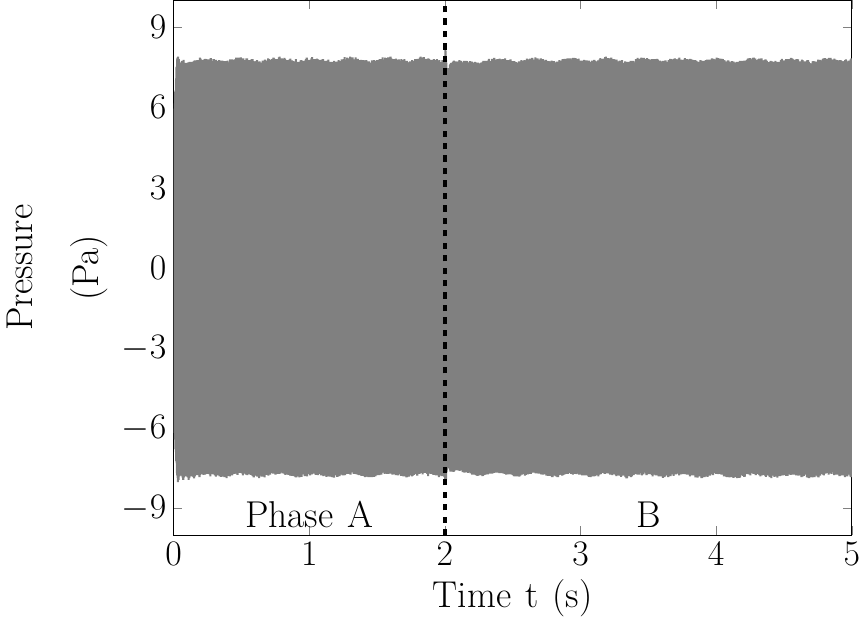}
		\subcaption{$a=6\ \text{Pa}$}\label{fig: pressure at a=6}
	\end{subfigure}
	\hfill
	\begin{subfigure}[t]{0.49\columnwidth}
		\includegraphics[width=\linewidth]{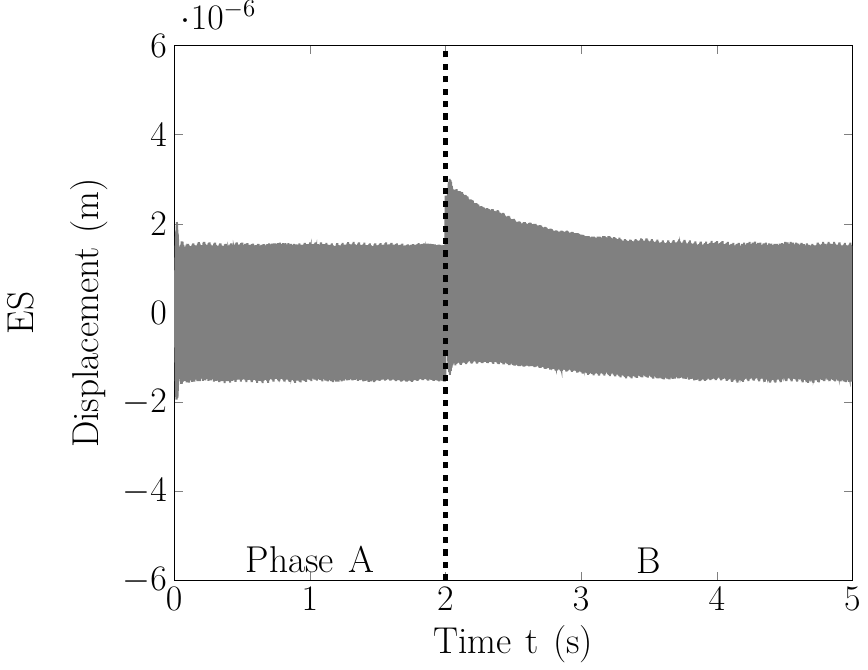}
		\subcaption{$a=6\ \text{Pa}$}\label{fig:ES displacement at a=6}
	\end{subfigure}\\
	
	\begin{subfigure}[t]{0.49\columnwidth}
		\includegraphics[width=\linewidth]{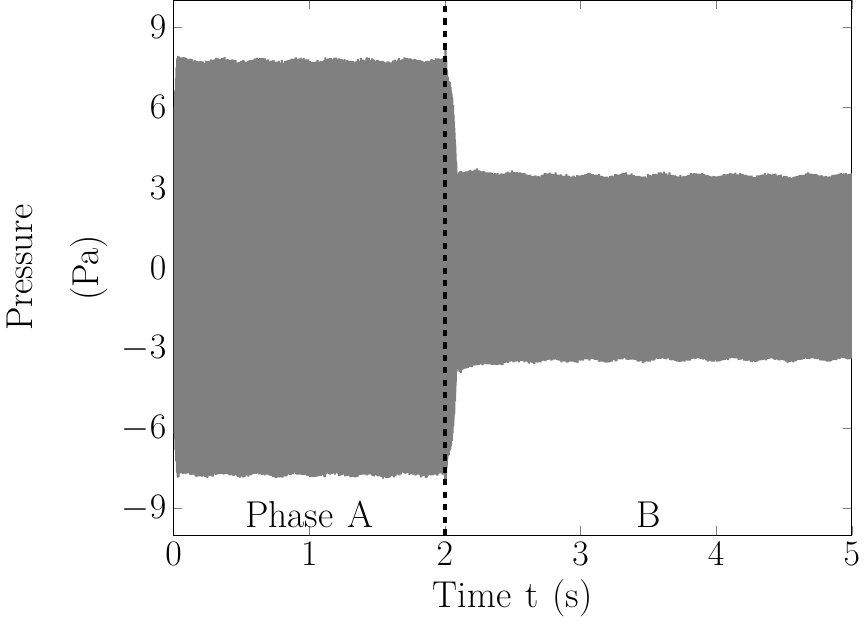}
		\subcaption{$a=8\ \text{Pa}$}\label{fig: pressure at a=8}
	\end{subfigure}
	\hfill
	\begin{subfigure}[t]{0.49\columnwidth}
		\includegraphics[width=\linewidth]{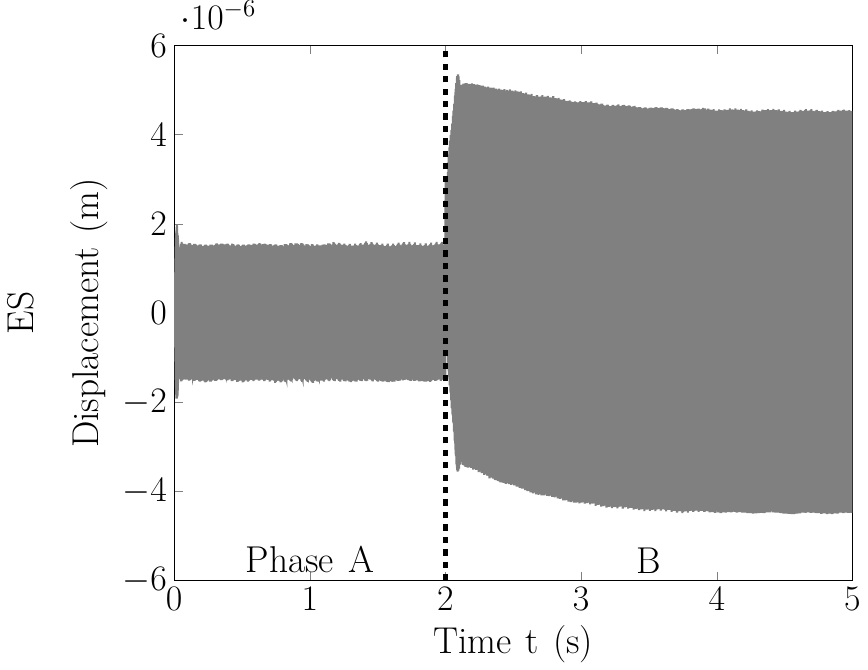}
		\subcaption{$a=8\ \text{Pa}$}\label{fig:ES displacement at a=8}
	\end{subfigure}\\
	
	\begin{subfigure}[t]{0.49\columnwidth}
		\includegraphics[width=\linewidth]{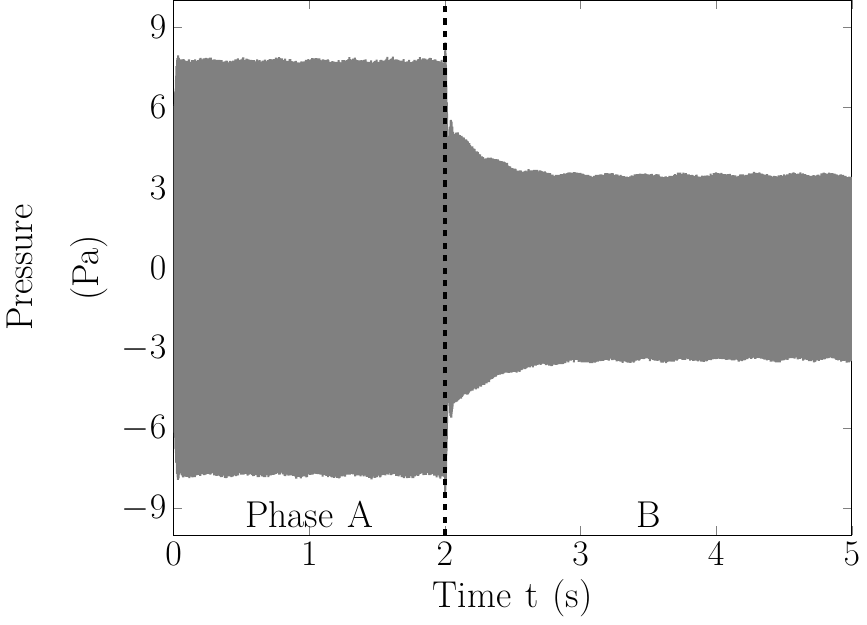}
		\subcaption{$a=15\ \text{Pa}$}\label{fig: pressure at a=15}
	\end{subfigure}
	\hfill
	\begin{subfigure}[t]{0.49\columnwidth}
		\includegraphics[width=\linewidth]{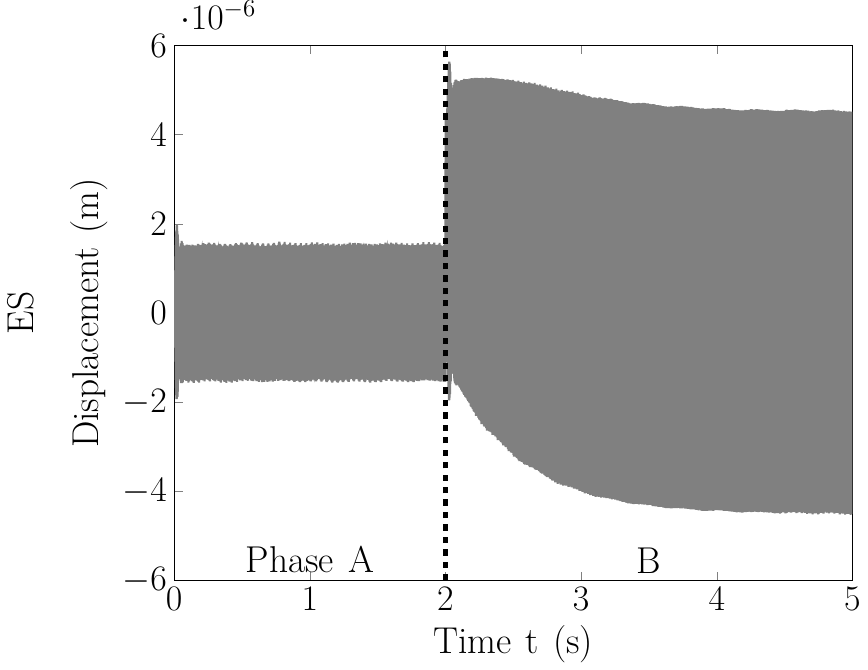}
		\subcaption{$a=15\ \text{Pa}$}\label{fig:ES displacement at a=15}
	\end{subfigure}
	\caption{Phases A and B of the measured pressure inside the reduced section tube (a,c,e), and estimated displacement of the ER's membrane computed in the control loop (b,d,f) for a 575 Hz sine excitation signal.}
	\label{fig:transient_ampl}
\end{figure}

The time histories of the measured pressure inside the $S_t$-tube and of the ES displacement are plotted in Fig. \ref{fig:transient_ampl} for different values of the $a=\{6,8,15\}\ \text{Pa}$, under a monoharmonic external sine excitation at 575 Hz. Fig.s \ref{fig: pressure at a=6} and \ref{fig:ES displacement at a=6} show that the value $a=6\ \text{Pa}$ is not sufficient to bring the ER's response to the higher energy equilibrium point, while for $a\geq8\ \text{Pa}$ the higher energy equilibrium point is reached, demonstrated by larger oscillation of the ES displacement, and lower amplitudes of the measured pressures in the $S_t$-tube in Fig.s \ref{fig: pressure at a=8}, \ref{fig:ES displacement at a=8}, \ref{fig: pressure at a=15}, \ref{fig:ES displacement at a=15}. As a result, the value of $a$ needed to achieve the desired equilibrium point under the initial conditions of the present experimental testbench, should lie somewhere between $a=6\ \text{Pa}$ and $a=8\ \text{Pa}$. Notice how the undesirable sudden increase of the ES displacement and of the measured pressure in the $S_t$-tube, at the beginning of phase B, is significantly weakened with respect to Fig. \ref{fig:pulse_p_coupled}, where $a$ was taken equal to $200$ Pa. Moreover, as expected, the duration of $f_{ext}^a$ increases with $a$. Therefore, by choosing the minimum value of $a$ needed to attain the desired equilibrium point, the transient duration can be minimized as well. On the other hand, a proper selection of the parameter $b$ appearing in Eq. \eqref{eq:form_fictive}, would also allow to minimize the transient duration for a fixed value of $a$. The optimization of the form of $f_{ext}^a$ is beyond the scope of the present paper and will be addressed in future publications.\\

The experimental tests reported in this section prove the potential of the proposed strategy to target desired equilibrium points of Duffing-like dynamics, at low excitation levels, in an acoustic environment made by enclosed cavities, where higher energy equilibrium points are required for efficient noise attenuation. Moreover, as known from \cite{de2022effect}, the placement of the ER in small acoustic cavities induces acoustic feedbacks on the ER which might jeopardize the stability of the system, whenever the ER acoustical passivity is lost. The absence of any stability issue during the experimental tests reported in this section, reinforce the reliability of the proposed concept. The investigation on the impact of an $f_{ext}^{ES}\neq f_{ext}^m$ upon stability is beyond the scope of the present paper, and will be addressed by a future dedicated study. 

\section{Conclusion}\label{sec:conclusions}
In this paper, we have conceived and validated a model-inversion feedforward control strategy able to synthesize specific equilibrium points of a desired nonlinear dynamics independently of the external excitation and for any initial condition. It is based upon the real-time integration of an exosystem dynamics described by a nonlinear Ordinary-Differential-Equation, where the measured external force is augmented by an additional purely artificial transient, which plays the role of forcing the plant to the desired equilibrium point. After having provided the general mathematical framework of the control concept, we have applied it for the synthesis of nonlinear Duffing-like generalized impedance of an Electroacoustic Resonator. By numerical simulations, we have proven that the additional synthetic external force allows to expand the basin of attraction of the target equilibrium point, therefore driving the exhibited solution towards the desired trajectory which is showcased by the plant also after the extinguishment of the artificial transient. Then, two experiments have been conducted to assess the performances of the control strategy applied to the Electroacoustic Resonator. The first experimental test is carried out in a quasi-open acoustic field environment and demonstrates the efficiency of the control concept to achieve desired equilibrium points of a Duffing-like target dynamics. The second one is carried out in an enclosed acoustic field, where the Electroacoustic Resonator is coupled to small acoustic cavities to challenge the robustness of the proposed control strategy when strong acoustic feedback might endanger the stability of the coupled system \cite{de2022effect}. In both experimental campaigns, the Electroacoustic Resonator achieved the desired equilibrium point under monoharmonic excitations, without featuring instability. Future studies will be dedicated to the optimization of shape, amplitude and duration of the designed artificial transient, taking into account also multi-harmonic excitations, or different types of nonlinearities. The choice of the value of the amplitude of the synthetic transient should be improved by the study of its influence on the basin of attraction using analytical tools, or selected in real-time with an adaptive law in future works. Other future developments of the proposed control strategy might involve a feedback architecture, contemplating several collocated microphones. After having enlarged the spectra of classical impedances towards the realm of synthetic nonlinear operators \cite{guo2020improving,DeBono2024,MORELL2024118437,morell2025experimental}, the work presented here is a fundamental step for the full controllability of digitally-programmed nonlinear impedances, further stimulating the scientific challenge of the inverse problem for the design of nonlinear optimal impedances for noise attenuation, in realistic scenarios. Thanks to the control strategy proposed here, indeed, the synthetic nonlinear behaviours implemented on Electroacoustic Resonators can finally find actual applications into realistic problem of noise control, such as for the attenuation of tonal noise in room acoustics. 

\section*{Acknowledgements}

The authors thank Abdelhakim Ezzerouki for providing helpful assistance in the electronic realization of the experiment. The authors would like to thank the following organizations for supporting this research: (i) The "Minist\`ere de la transition \'ecologique" and (ii) LABEX CELYA (ANR-10-LABX-0060) of the "Universit\'e de Lyon" within the program "Investissement d\text{'}Avenir" (ANR-11- IDEX-0007) operated by the French National Research Agency (ANR).\\
  \bibliographystyle{elsarticle-num} 
  \bibliography{References_These}



%
%
%
\end{document}